\documentclass{article}

\usepackage{PRIMEarxiv}

\usepackage[utf8]{inputenc} % allow utf-8 input
\usepackage[T1]{fontenc}    % use 8-bit T1 fonts
\usepackage{hyperref}       % hyperlinks
\usepackage{url}            % simple URL typesetting
\usepackage{booktabs}       % professional-quality tables
\usepackage{amsfonts}       % blackboard math symbols
\usepackage{nicefrac}       % compact symbols for 1/2, etc.
\usepackage{microtype}      % microtypography
\usepackage{graphicx}

\usepackage{amsmath,amsfonts,amssymb,amsthm}
\usepackage{physics}
\usepackage{tikz,pgfplots}
\usepackage{dirtytalk}
\usepackage{subcaption}
\newtheorem{definition}{Definition}

\newcommand\NUM{\addtocounter{equation}{1}\tag{\theequation}}

\def\mean#1{\left\langle #1 \right\rangle}
\newcommand\D{\mathrm{d}}
\newcommand\dt{\D t \;}

\newcommand{\fx}{{\mathcal{X}}}
\newcommand{\fp}{{\mathcal{P}}}

\newcommand{\vx}{{\pmb{x}}}
\newcommand{\vxi}{{\pmb{\xi}}}

\newcommand\xp{x_\mathrm{P}}
\newcommand\xf{x_\mathrm{F}}
\newcommand\xisi{x_\mathrm{ISI}}
\newcommand\tp{{t_\mathrm{P}}}
\newcommand\tf{{t_\mathrm{F}}}
\newcommand\dtp{\D \tp}
\newcommand\dtf{\D \tf}
\newcommand\one{^{(1)}}
\newcommand\two{^{(2)}}

\title{An Information-Geometric Formulation of Pattern Separation and Evaluation of Existing Indices
}

\author{
  Harvey Wang\\
  Faculty of Computer Science\\
  Dalhousie University\\
  Halifax, Nova Scotia, Canada\\
  \texttt{harvey@dal.ca} \\
   \And
  Selena Singh\\
  Department of Psychology, Neuroscience \& Behaviour\\
  McMaster University\\
  Hamilton, Ontario, Canada\\
  \texttt{singhs11@mcmaster.ca} \\
   \And
  Thomas Trappenberg\\
  Faculty of Computer Science\\
  Dalhousie University\\
  Halifax, Nova Scotia, Canada\\
  \texttt{tt@cs.dal.ca} \\
   \And
  Abraham Nunes\\
  Faculty of Computer Science and Department of Psychiatry\\
  Dalhousie University\\
  Halifax, Nova Scotia, Canada\\
  \texttt{nunes@dal.ca} \\
}

\begin{document}
\maketitle

\begin{abstract}
Pattern separation is a computational process by which dissimilar neural patterns are generated from similar input patterns. We present an information-geometric formulation of pattern separation, where a pattern separator is modelled as a family of statistical distributions on a manifold. Such a manifold maps an input (i.e. coordinates) to a probability distribution that generates firing patterns. Pattern separation occurs when small coordinate changes result in large distances between samples from the corresponding distributions. Under this formulation, we implement a two-neuron system whose probability law forms a 3-dimensional manifold with mutually orthogonal coordinates representing the neurons’ marginal and correlational firing rates. We use this highly controlled system to examine the behaviour of spike train similarity indices commonly used in pattern separation research. We found that all indices (except scaling factor) were sensitive to relative differences in marginal firing rates, but no index adequately captured differences in spike trains that resulted from altering the correlation in activity between the two neurons. That is, existing pattern separation metrics appear (A) sensitive to patterns that are encoded by different neurons, but (B) insensitive to patterns that differ only in relative spike timing (e.g. synchrony between neurons in the ensemble).
\end{abstract}

% keywords can be removed
\keywords{pattern separation \and information geometry \and information theory}

%===============================================================================
% INTRO
%===============================================================================
\section{Introduction}

\subsection{Pattern Separation in Computational Neuroscience}

The hippocampus plays a key role in the formation of complex associative and episodic memories \cite{tulving_episodic_1998, scoville_loss_1957}. Classical computational models have proposed that the hippocampus performs two complementary neural computations to minimize interference and maximize information storage: \textit{pattern separation} and \textit{pattern completion} \cite{mcclelland_why_1995}. Pattern separation is a computation performed by a neural network that minimizes the similarity between distinct but overlapping input patterns \cite{wigstrom_model_1974}. It is thought to be performed prior to long-term memory storage to reduce the probability of interference in memory recall and enhance downstream pattern completion \cite{rolls_computational_2006}. Pattern completion, in contrast to pattern separation, is performed by a network during the retrieval of stored patterns when presented with partial or degraded input patterns.

The hippocampal dentate gyrus has a set of properties that makes it ideally suited to perform pattern separation, such as the sparse and competitive firing of granule cells \cite{espinoza_parvalbumin_2018, guzman_how_2021}. Rodent studies have shown that activity patterns in the dentate gyrus are less correlated than those in the entorhinal cortex and hippocampal CA3 region, consistent with dentate gyrus pattern separation \cite{neunuebel_ca3_2014}. Importantly, pattern separation is found not only in the hippocampus. The cerebellum and insect mushroom body also have properties that would facilitate pattern separation \cite{cayco-gajic_re-evaluating_2019}. Thus, pattern separation is a fundamental computation utilized by the brain to reduce interference between activity patterns to promote functions such as memory encoding.

Physiological changes in the dentate gyrus have been linked to conditions involving cognitive impairments, including schizophrenia \cite{nakahara_dentate_2020}, Alzheimer's disease \cite{ohm_dentate_2007}, and epilepsy \cite{young_upregulation_2009}. Patients affected by these conditions present with reduced abilities to distinguish between similar stimuli \cite{das_loss_2014, parizkova_spatial_2020, madar_deficits_2021}. There is, therefore, great interest in detailed biophysical mechanisms underlying pattern separation. In particular, many computational models have been designed to simulate the neurophysiological circuitry responsible for pattern separation under various physiological conditions \cite{myers_role_2009, chavlis_pattern_2017, santhakumar_role_2005, madar_pattern_2019}. However, when studied in isolation, \say{pattern separation} is often defined even more arbitrarily, often reduced to refer to any computation that minimizes similarity or overlap (see further discussions in Sections \ref{sec:no_single_def} and \ref{sec:existing_indices}), and many of the aforementioned studies often employ inconsistent measurements for it. How do we simulate a computation for which there is no precise definition? Such definition, therefore, is essential to further the understanding of what pattern separation is, and is not, as well as providing a theoretical basis for further studies.

\subsection{There Exists no Single Definition for Pattern Separation}
\label{sec:no_single_def}

Pattern separation is generally defined as a computation that maximizes \textit{dissimilarity} or \textit{orthogonalization} \cite{santoro_reassessing_2013, mcnaughton_hebb-marr_1990} between neural patterns, given similar yet distinct input patterns \cite{wigstrom_spatial_1977, rolls_mechanisms_2013, yassa_pattern_2011}. Definition \ref{def:classical_ps} formalizes this concept. 

\begin{definition}
\label{def:classical_ps}
For some dissimilarity or distance measure $d$ and input neural patterns $\fx^{(1)}$ and $\fx^{(2)}$, the mapping $f$ performs pattern separation if and only if the dissimilarity between its outputs are greater than that between $\fx^{(1)}$ and $\fx^{(2)}$,
$$
    d\left(\fx^{(1)}, \fx^{(2)}\right)
    < d\left(
        f\left(\fx^{(1)}\right), f\left(\fx^{(2)}\right)
    \right).
$$
\end{definition}

Clearly, the definition of pattern separation relies heavily on the form of the dissimilarity measure $d$. Definition \ref{def:classical_ps} is in fact not a definition for a single, but rather, for multiple computations. This creates a problem where different forms of $d$ would translate to different interpretations for pattern separation. For example, if Pearson correlation is used for $d$, then pattern separation would be a computation that minimizes \textit{correlation}; if $d$ is the Hamming distance, then pattern separation is a computation that minimizes \textit{overlap}.

Having a diverse set of indices, all describing ``pattern separation", limits comparability of pattern separation efficacy across computational studies \cite{santoro_reassessing_2013, vineyard_quantifying_2016, chavlis_pattern_2017}. \cite{madar_temporal_2019} showed that quantifying pattern separation abilities of a system by applying different measures, $d$, to the same dataset lead to different conclusions; for example, they showed that pairs of spike trains can be uncorrelated without being orthogonal, and therefore suggested that pattern separation should be considered \textit{a group of potential computations}, since it is unclear what spike train features are most relevant to the brain and what constitutes similarity or dissimilarity. For instance, if information was coded using a neuron's firing rate, a pattern separator would aim to convert a series of input trains with similar rates to one with dissimilar rates. On the other hand, the firing rate could carry no information, and the system would rather use spike timing (or temporal coding) to represent information; in this case, the pattern separator may aim to simply change the relative timing of spike trains. A recent analysis of the most commonly used pattern separation indices demonstrated that some failed to capture information loss that may occur with high degrees of sparsity (a property often associated with strong pattern separation), and instead found that these measures confounded information loss with excellent pattern separation \cite{bird_robust_2024}. Although these studies evaluated existing pattern separation measures, the evaluations were not conducted in such a way that would control for degrees of separation achieved by different pattern separation strategies. Taken together, these analyses of the most commonly used pattern separation measures highlight major issues with consistency, measurement accuracy, and applicability between studies, and highlight the importance of developing a single, unifying definition of pattern separation, along with a controlled re-evaluation of existing measures.

\subsection{An Information-Geometric Formulation of Pattern Separation}

In this work, we use tools from information geometry to describe pattern separation as a \textit{single} computation. This formulation is consistent with the existing notion of pattern separation (Definition \ref{def:classical_ps}), but also addresses shortcomings of existing definitions and measures.

A pattern separator (e.g. a neural network such as the hippocampal dentate gyrus) and its inputs (e.g. from the entorhinal cortex) are respectively modelled as a statistical manifold and its coordinates. A point on this manifold is a probability distribution that (A) is identified by its input coordinates and (B) generates output patterns (i.e., spike trains) as samples. A highly effective separator maps input patterns with small differences to output patterns with large differences. That is, small difference between input spike trains representing different patterns (which here constitute small changes in coordinates), would result in very different output spike trains from the neural pattern separator. Under our formulation, an excellent pattern separator would correspond to a statistical manifold with high curvature that maps small changes in its coordinates to distributions whose samples are highly distant or distinct from each other. In such manifolds, a single sample from a distribution provides a large degree of information about the underlying parameters or inputs by which it was generated. This is not only consistent with traditional notions of pattern separation, but also satisfies the problem identified by \cite{bird_robust_2024}, where common measures can fail to distinguish between pattern separation and information loss (which could occur by randomly adding noise to a pattern).

We demonstrate an application of this formalism by implementing a two-neuron system with an analytically tractable probability law, such that distances between spike-train distributions on the underlying manifold can be controlled. Using this system, we can simulate pattern separation that occurs by (A) changing \textit{which neurons} are active across patterns, (B) changing \textit{how much} the neurons are active, and (C) changing the \textit{relative timing} of neural spikes across patterns. We then evaluate the degree to which several commonly used pattern separation indices are sensitive to each of these changes.

\subsection{An Evaluation of Existing Indices of Pattern Separation}
\label{sec:intro_eval}

Existing indices appeared much more sensitive to pattern separation as a result of changes in the neuron's firing rates, and appear, in general, insensitive to pattern separation as a result of changes in the rate of the neuron's coincident firing (and thus the relatively timing by which the neurons are active). In addition, our results provide further evidence that, while sensitive to changes in neuronal firing rates, the variety of existing of pattern separation could lead to inconsistent conclusions on the amount of pattern separation performed.

%===============================================================================
% Section 2
%===============================================================================
\section{An Information-Geometric Formulation of Pattern Separation}

\subsection{Pattern Separator Described by a Statistical Manifold}
\label{sec:ps_manifold}

Let us use a $N$-dimensional statistical manifold $\mathcal{P}$ to describe the neural activity of a pattern separator, such as the hippocampal dentate gyrus, in response to different inputs. Consider the manifold's coordinates $\vxi$, representing parameters that influence the separator's outputs. These parameters can be subdivided into two different types: (A) those that are intrinsic to the system, such as network connectivity and other biophysical properties, and (B) those received as input patterns from other areas, such as the perforant path inputs from entorhinal cortex to the hippocampal dentate gyrus. Each distinct parameterization $\pmb{\xi}$ specifies a probability distribution $p(\vx; \vxi)$ on $\fp$, from which neural patterns $\vx$ may be drawn. In the case that the intrinsic parameters are fixed, the change in coordinates $\mathrm{d}\pmb{\xi} = (\mathrm{d}\xi_i)_{i=1, \dots, N}$ describes the change in the inputs, and the distance between the probability distributions reflects the change in the outputs.

A highly effective pattern separator maps input patterns with small differences to output patterns with large differences. Under our information-geometric formulation, this would correspond to a manifold $\fp$ with high curvature, which maps small changes in its coordinates  $\D\vxi$, to distributions whose samples are highly distant or distinct from each other, $p(\vx; \vxi)$ and $p(\vx; \vxi + \D\vxi)$. Under this model, a single output sample provides a large degree of information about the underlying parameters by which it was generated.

The definition of distance between probability distributions $p(\pmb{x}; \pmb{\xi})$ and $p(\pmb{x}; \pmb{\xi} + \mathrm{d}\pmb{\xi})$ is given by the \textit{Fisher information matrix}. The direction and magnitude of change in a probability distribution, as a result of a small change $\mathrm{d} \xi_i$ in the $i^\mathrm{th}$ coordinate, are described by the derivative with respect to $\mathrm{d}\xi_i$,
$$\mathcal{L}(\xi_i)=\pdv{\log p(\pmb{x}; \pmb{\xi})}{\xi_i},$$
where $\mathcal{L}$ is known as the score function. Elements of the Fisher information matrix $G=\left(g(\xi_i, \xi_j)\right)_{i,j=1,\dots,N}$ are given by the expectation of the product of the score functions for two coordinates
\begin{align*}
    g(\xi_i, \xi_j)
    &= \mean{ \mathcal{L}(\xi_i) \mathcal{L}(\xi_j)}_{\pmb{x}}
    \\&= \mean{
            \pdv{\log p(\pmb{x}; {\pmb{\xi}})}{\xi_i}
            \pdv{\log p(\pmb{x}; {\pmb{\xi}})}{\xi_j}
        }_{\pmb{x}},
    \NUM
    \label{eq:fisher_matrix}
\end{align*}
where $\mean{f(\vx)}_{\vx}$ denotes the expectation value of a function $f(\vx)$, defined as the weighted average over all possible values of $\vx$, $\sum_\vx f(\vx) p(\vx)$. The squared distance between two probability distributions defined by $\vxi$ and $\vxi + \D\vxi$ can then be computed by the quadratic form of $\mathrm{d} {\pmb{\xi}}$,
\begin{equation}
    \mathrm{d} s^2
    = \sum_{i=1}^{N} \sum_{j=1}^{N}
        g(\xi_i, \xi_j)\; \mathrm{d} \xi_i \mathrm{d}\xi_j.
    \label{eq:sq_dist_n}
\end{equation}

In our formulation of pattern separation, two different coordinates, $\vxi$ and $\vxi + \D\vxi$, represent two different input patterns. These coordinates define two different probability distributions, $p(\pmb{x}; \pmb{\xi})$ and $p(\pmb{x}; \pmb{\xi} + \mathrm{d} \pmb{\xi})$, whose samples are less overlapping, or more separated, as a result of $\D\vxi$. This conceptualization of pattern separation is consistent with its existing notion given in Definition \ref{def:classical_ps}, and provides a clear mathematical formalism that we demonstrate when analyzing different pattern separation metrics.

It is important to note that while we can \textit{directly} simulate changes $\mathrm{d}\pmb{\xi}$ theoretically, $\pmb{\xi}$ and $\mathrm{d}\pmb{\xi}$ are only \textit{indirectly} influenced in real neural systems. For example, pattern separation may be achieved by drastically altering intrinsic parameters such as network connectivity and neuronal biophysics. We emphasize here that these properties may change within biological systems and affect pattern separation performance, but these changes occur over longer time scales than what is most relevant for pattern separation of two similar and highly overlapping stimuli presented consecutively. We, however, still consider these mechanisms of achieving pattern separation. As such, to evaluate existing indices we manipulate $\mathrm{d}\pmb{\xi}$ to simulate changes in input patterns or physiological properties of the network, giving us a ground truth for the \textit{type} and \textit{degree} of pattern separation simulated.
% by the network's inputs (as well as possibly physiological changes in the neural system)

%In general, the larger the magnitudes of $\mathrm{d} \pmb{\xi}$, the larger the distance between $p(\pmb{x}; \pmb{\xi})$ and $p(\pmb{x}; \pmb{\xi}+\mathrm{d}\pmb{\xi})$, and the larger the difference between samples $\pmb{x}^{(1)} \sim p(\pmb{x}; \pmb{\xi})$ and $\pmb{x}^{(2)} \sim p(\pmb{x}; \pmb{\xi} + \mathrm{d} \pmb{\xi})$. This is particularly true in coordinate systems that contain orthogonal coordinates, where changes in each coordinate can independently contribute to the increase in distance.

%In our evaluation of existing indices of pattern separation, we postulate that an appropriate index should be sensitive to the increase in distance between $p(\pmb{x}; \pmb{\xi})$ and $p(\pmb{x}; \pmb{\xi} + \mathrm{d} \pmb{\xi})$ caused by $\mathrm{d} \pmb{\xi}$.

\subsection{A Two-Neuron Manifold and Its Orthogonal Coordinates}
\label{sec:2_neuron}

Consider a model of a system of two neurons \cite{nakahara_information-geometric_2002} and its time-independent firing pattern $\pmb{x}=\left(x_i\right)_{i=1,2}$ in a given observation window, where $x_i=1$ and $x_i = 0$ indicate that the $i$th neuron is active (with one or more spikes) or silent, respectively. We use the highly controlled probability distribution $p(\vx; \vxi)$ for this model to separately alter the independent firing rates and the degree of correlation between neurons. Under the formalism introduced above, these manipulations change the coordinates on the system's manifold. Controlling the amount of pattern separation, as well as the strategy by which it is achieved, facilitated our evaluation of existing indices. Specifically, we can use this model to simulate pattern separation that arises from changes in each neuron's independent firing rates, relative timing that they are active, \textit{or} both.

The probability distribution $p(\pmb{x}; \pmb{\xi})$ for this model is given by the probabilities that each neuron is active or silent,
$$
    q_{mn} = \mathrm{Prob}\{x_1 = m,\; x_2 = n\},\qquad m,n=0,1.
$$
These probabilities give an exact expansion of the logarithm of the probability distribution of the system,
\begin{equation*}
    \log p(\pmb{x}; \vxi) = \log \frac{q_{10}}{q_{00}} x_1 + \log \frac{q_{01}}{q_{00}} x_2 + \theta x_1 x_2 + \log q_{00},
\end{equation*}
where
\begin{equation}
    \theta = \log \frac{q_{11}q_{00}}{q_{10}q_{01}}.
\end{equation}
The term $\theta$ specifies the within-ensemble correlation that represents the degree to which neurons 1 and 2 are correlated. $\theta$ is $0$ when the two neurons are uncorrelated, approaches $-\infty$ as the two neurons become \textit{maximally anti-correlated} (as $q_{11}$ and $q_{00}$ approach $0$), and approaches $\infty$ as the two neurons become \textit{maximally correlated} (as $q_{10}$ and $q_{01}$ approach $1$).

Among the set of four probabilities, $\{q_{00}, q_{10}, q_{01}, q_{11}\}$, only three variables are free, due to the constraint $q_{00} + q_{10} + q_{01} + q_{11} = 1.$ The manifold for this two-neuron model is therefore three-dimensional. Though there are many coordinate systems for this model (for example, any of the three from $\{q_{00}, q_{10}, q_{01}, q_{11}\}$), we use
$$\pmb{\xi}=(\eta_1, \eta_2, \theta)$$
to simulate different types of pattern separation, where the variables
\begin{align*}
    \eta_1 = \mean{x_1} = q_{10} + q_{11},
    \qquad
    \eta_2 = \mean{x_2} = q_{01} + q_{11},
    \NUM
\end{align*}
denote the marginal firing rates of neurons 1 and 2, respectively. This is a convenient coordinate system because $(\eta_1, \eta_2)$ and $\theta$ are mutually orthogonal, meaning that their directions of change are uncorrelated \cite{nakahara_information-geometric_2002}. Specifically, this means that
\begin{equation}
    g(\eta_1, \theta) = g(\eta_2, \theta) = 0.
\end{equation}

Crucially, the orthogonality between $(\eta_1, \eta_2)$ and $\theta$ allows us to increase the distance between two probability distributions by changing \textit{only} $\mathrm{d}\theta$ \textit{or} one of $\mathrm{d}\eta_1$ and $\mathrm{d}\eta_2$. In other words, for two distributions with the same values of $\eta_1$ and $\eta_2$, the distance between them is proportional to the difference between their $\theta$ values,
\begin{equation}
    \mathrm{d} s^2 \Big|_{\mathrm{d} \eta_1 = \mathrm{d}\eta_2 = 0} =  g(\theta, \theta) \; (\mathrm{d} \theta)^2.
    \label{eq:sq_dist_theta}
\end{equation}
Similarly, this orthogonality simplifies Eq. \ref{eq:sq_dist_n} in the case where $\mathrm{d}\theta=0$
\begin{equation}
    \mathrm{d} s^2 \Big|_{\mathrm{d} \theta = 0}
        = 2\; g(\eta_1, \eta_2) \; \mathrm{d}\eta_1 \mathrm{d}\eta_2
            + g(\eta_1, \eta_1) \; (\mathrm{d} \eta_1)^2
            + g(\eta_2, \eta_2) \; (\mathrm{d} \eta_2)^2 .
    \label{eq:sq_dist_eta2_dtheta_zero}
\end{equation}

Using this setup, we define a \say{control} system by specifying its probability distribution using these three variables, denoted $p(\pmb{x}; \eta_1, \eta_2, \theta)$. Using similar notation, we define a second \say{comparison} distribution $p(\pmb{x}; \eta_1 + \mathrm{d}\eta_1, \eta_2 + \mathrm{d}\eta_2, \theta + \mathrm{d}\theta)$, and thus simulate pattern separation by systematically manipulating the values of $\mathrm{d}\eta_1$, $\mathrm{d}\eta_2$, and $\mathrm{d}\theta$ to alter the distance between $p(\pmb{x}; \eta_1, \eta_2, \theta)$ and $p(\pmb{x}; \eta_1 + \mathrm{d}\eta_1, \eta_2 + \mathrm{d}\eta_2, \theta + \mathrm{d}\theta)$. The changes in parameters of the two distributions (i.e., $\mathrm{d}\eta_1$, $\mathrm{d}\eta_2$, and $\mathrm{d}\theta$) represent the differences between two input patterns presented to a pattern separator, and the distance between the distributions represent the differences between corresponding outputs from the separator. Pattern separation is therefore achieved in this system when small $\mathrm{d}\eta_1$, $\mathrm{d}\eta_2$, and $\mathrm{d}\theta$ result in a large distance between the control and comparison distributions.

%The $i^\mathrm{th}$ and $j^\mathrm{th}$ coordinates on the manifold representing the pattern separator are said to be \textit{orthogonal} when their directions of change are uncorrelated \cite{nakahara_information-geometric_2002}. In other words, coordinates $\xi_i$ and $\xi_j$ are orthogonal if $g(\xi_i, \xi_j)=0$.

An appropriate index of pattern separation should be sensitive to the increased distance between $p(\pmb{x}; \eta_1, \eta_2, \theta)$ and $p(\pmb{x}; \eta_1 + \mathrm{d}\eta_1, \eta_2 + \mathrm{d}\eta_2, \theta + \mathrm{d}\theta)$. This increased distance corresponds to pattern separation as a result of changes in each neuron's firing rates, relative timing of activity, \textit{or} a combination of both.

\subsection{Generating Diverging Patterns}
\label{sec:diverge_patterns}

Neural patterns were generated from control and comparison distributions $p(\pmb{x}; \eta_1, \eta_2, \theta)$ and $p(\pmb{x}; \eta_1 + \mathrm{d}\eta_1, \eta_2 + \mathrm{d}\eta_2, \theta + \mathrm{d}\theta)$, by sampling $N_\mathrm{bins}=10^3$ times from each distribution. Each sample constituted a \say{time bin} in which one, both, or neither of the neurons may be active; that is, samples were drawn from categorical distributions with event probabilities $\{q_{00}, q_{10}, q_{01}, q_{11}\}$ for each time bin.

\subsubsection{Patterns with Dissimilar Neuronal Coincident Firing Rates}
\label{sec:diverge_theta}

To assess the sensitivity of existing indices in detecting pattern separation as a result of changes in within-ensemble, neuron-neuron correlation structures, we set the marginal firing rates of the two neurons constant and gradually increased $\mathrm{d}\theta$. Specifically, we set the control distribution to $p(\pmb{x}; \eta_1, \eta_2, 0)$, and the comparison to $p(\pmb{x}; \eta_1, \eta_2, \mathrm{d}\theta)$. Recall from Eq. \ref{eq:sq_dist_theta}, the squared distance between the two distributions is proportional to $\left(\mathrm{d}\theta\right)^2$ in this setup.

We assigned new values for $\theta$ by changing $q_{11}$ and $q_{00}$ the same amount in the same direction, while changing $q_{10}$ and $q_{01}$ the same amount in the opposite direction, thereby keeping $\eta_1$ and $\eta_2$ constant. In other words, $q_{00}, q_{10}, q_{01}, \text{and } q_{11}$ were updated by the following
\begin{align*}
    q_{00} \leftarrow q_{00} + \mathrm{d} q_{11}, \quad
    q_{10} \leftarrow q_{10} - \mathrm{d} q_{11}, \quad
    q_{01} \leftarrow q_{01} - \mathrm{d} q_{11}, \quad
    q_{11} \leftarrow q_{11} + \mathrm{d} q_{11}, \quad
\end{align*}
so that the corresponding new $\theta$ was
\begin{align*}
    \theta \leftarrow \log \frac
        {\left(q_{11}+\mathrm{d} q_{11}\right) \left(q_{00} +\mathrm{d} q_{11}\right)}
        {\left(q_{10}-\mathrm{d} q_{11}\right) \left(q_{01} -\mathrm{d} q_{11}\right)},
\end{align*}
where $\eta_1$ and $\eta_2$ remained the same
\begin{align*}
    \eta_1 \leftarrow
        \left( q_{10} - \mathrm{d} q_{11} \right)
        + \left( q_{11} + \mathrm{d} q_{11} \right)
        = q_{10} + q_{11},
    \\
    \eta_2 \leftarrow \left( q_{01} - \mathrm{d} q_{11} \right)
        + \left( q_{11} + \mathrm{d} q_{11} \right)
        = q_{01} + q_{11}.
\end{align*}

This procedure was repeated for different values of $\eta_1$ and $\eta_2 \ge \eta_1$. We also applied the additional constrains $\eta_2=1-\eta_1$ to keep the overall firing rate of the system constant.

Overall, this experimental setup created comparison distributions whose two neurons ranged from nearly maximally anti-correlated (when $q_{11} = q_{00} = 10^{-2}$ so that $\mathrm{d}\theta$ was small) to nearly maximally correlated (when $q_{10} = 10^{-2}$ so that $\mathrm{d}\theta$ was large).

\subsubsection{Patterns with Dissimilar Neuronal Firing Rates}
\label{sec:diverge_eta2}
To assess the sensitivity of existing indices in detecting pattern separation as a result of changes in one neuron's firing rate, we set the marginal firing rate of one of the neurons constant, allowed the neurons to fire independently, and gradually increased $\mathrm{d}\eta_2$. Specifically, we set the control distribution to $p(\pmb{x}; \eta_1, 0.1, 0)$, and the comparison to $p(\pmb{x}; \eta_1, 0.1+\mathrm{d}\eta_2, 0)$. The squared distance between these two distributions is proportional to $\left(\mathrm{d}\eta_2\right)^2$:
\begin{equation}
    \mathrm{d} s^2 \Big|_{\mathrm{d}\eta_1 = \mathrm{d} \theta = 0}
        = g(\eta_2, \eta_2) \; (\mathrm{d} \eta_2)^2 .
    \label{eq:sq_dist_eta2}
\end{equation}

We defined our systems by expressing $\{q_{00}, q_{10}, q_{01}, q_{11}\}$ as functions of $\eta_1$ and $\eta_2$ at $\theta = 0$. Since $\theta = 0$ means that neurons 1 and 2 fire independently, $q_{11} = \eta_1\eta_2$. We could also, of course, show this with directly using the two-neuron model, with
\begin{align*}
    \theta = \log \frac{q_{11}q_{00}}{q_{10}q_{01}} = 0 &\iff \frac{q_{11}q_{00}}{q_{10}q_{01}} = 1
    %\\&\iff q_{11}q_{00} = q_{10}q_{01}.
    \iff q_{11} = \frac{q_{10}q_{01}}{q_{00}}.
\end{align*}
Since $q_{10} = \eta_1 - q_{11}$, $q_{01} = \eta_2 - q_{11}$, and $q_{00} = 1 - q_{10} - q_{01} - q_{11}$, we can rearrange the above,
\begin{align*}
    q_{11}
        &= \frac{(\eta_1 - q_{11})(\eta_2 - q_{11})}{1 - (\eta_1 - q_{11}) - (\eta_2 - q_{11}) - q_{11}}
        %= \frac{q_{11}(\eta_1 - 1)(\eta_2 - 1)}{1 - \eta_1 + q_{11} - \eta_2 + q_{11} - q_{11}}
    \\&\iff (\eta_1 - 1)(\eta_2 - 1) = 1 - \eta_1 - \eta_2 + q_{11}
    %\\&\iff \eta_1\eta_2 - \eta_1 - \eta_2 + 1 = 1 - \eta_1 - \eta_2 + q_{11}
    \\&\iff q_{11} = \eta_1\eta_2.
\end{align*}

Overall, this experimental setup created comparison distributions that permitted one of the neurons to range from sparsely active (when $\mathrm{d}\eta_2$ was small) to densely firing (when $\mathrm{d}\eta_2$ was large).

%===============================================================================
% EXISTING INDICES
%===============================================================================
\section{Existing Indices}
\label{sec:existing_indices}

Recall from Definition \ref{def:classical_ps}, pattern separation is characterized by the increase of pairwise dissimilarity between two neural patterns. Specifically, the mapping $f$ performs pattern separation if and only if,
$$
    d\left(\fx^{(1)}, \fx^{(2)}\right)
        < d\left(
            f\left(\fx^{(1)}\right), f\left(\fx^{(2)}\right)
        \right),
$$
where $d$ is some index of dissimilarity between two neural patterns $\fx^{(1)}$ and $\fx^{(2)}$. The superscripts (1) and (2) respectively denote the first and second pattern examined. Individual indices are denoted $d$ with subscripts.

\subsection{Pattern Representation}
\label{sec:pat_rep}

Existing indices of pattern separation operate operate on one of two representations of neural patterns: discretized vectors $\vx$ or non-discretized lists $T$ (the notation $\fx$ above merely denotes pattern as a random variable, but makes no specification on how the pattern is represented).

\subsubsection{Non-discretized Patterns}

Consider a system of $N_{\mathrm{neurons}}$ neurons. The spiking pattern of this system can be represented as a list of spike trains from each individual neuron, $T = \{T_{i}\}_{i=1, \dots, N_\mathrm{neurons}}$. $T_i = \{T_{ij}\}_{j = 1, \dots, n_i}$ denotes the list of spike times $T_{ij}$ from the $i^\mathrm{th}$ neuron, where and $n_i$ is the total number of times the neuron fires.

\subsubsection{Discretized Patterns}

A pattern $T$ can be binned and binarized, so that a wider variety of indices can be applied. This is generally done by dividing the observation window into $N_\mathrm{bins}$ equally sized bins, and assigning each bin a binary value $x_{ij} \in \{0, 1\}$. $x_{ij} = 1$ indicates that the $i^\mathrm{th}$ neuron is active at least once in the $j^\mathrm{th}$ bin, and  $x_{ij} = 0$ otherwise. The matrix representation of a binarized pattern is then simply $X = (x_{ij})_{i=1, \dots, N_\mathrm{neurons}}^{j = 1, \dots, N_{\mathrm{bins}}}$. These matrices are then generally vectorized into
\begin{align*}
    \pmb{x}
       &= \mathrm{vec}({X^\intercal})\\
       %&= [\pmb{x}_{1}, \pmb{x}_{2}, \dots, \pmb{x}_{N}] \\
       &= [
            x_{11}, x_{12}, \dots, x_{1N_{\mathrm{bins}}},
            x_{21}, x_{22}, \dots, x_{2N_{\mathrm{bins}}}, \dots
            x_{N_\mathrm{neurons}1}, x_{N_\mathrm{neurons}2}, \dots, x_{N_\mathrm{neurons}N_\mathrm{bins}}
            ].
        \NUM
    \label{eq:binned_vector}
\end{align*}

\subsection{Common Indices}
\label{sec:common-indices}

The \textbf{Pearson correlation},
\begin{equation}
    d_\rho\left(\pmb{x}^{(1)}, \pmb{x}^{(2)}\right) =
        \frac{
            \mean{\pmb{x}^{(1)}\pmb{x}^{(2)}}
            - \mean{\pmb{x}^{(1)}}\mean{\pmb{x}^{(2)}}
        } {
            \sqrt{
                \mean{\left(\pmb{x}^{(1)}\right)^2}
                - \mean{\pmb{x}^{(1)}}^2
            }
            \sqrt{\mean{\left(\pmb{x}^{(2)}\right)^2} - \mean{\pmb{x}^{(2)}}^2}
        },
    \label{eq:pearson}
\end{equation}
is one of the most widely used indices. It has been used to study pattern separation both in computational \cite{yim_intrinsic_2015} and experimental studies \cite{neunuebel_ca3_2014, leutgeb_pattern_2007, senzai_physiological_2017, danielson_vivo_2017}, as well as under various physiological conditions, such as during epileptic hyperexcitability in the dentate gyrus \cite{yim_intrinsic_2015}.

The \textbf{cosine similarity}, or the normalized dot product,
\begin{equation}
    d_\theta\left(\pmb{x}^{(1)}, \pmb{x}^{(2)}\right)
        = \frac
            {\pmb{x}^{(1)} \cdot \pmb{x}^{(2)}}
            {\big\|\pmb{x}^{(1)}\big\| \big\|\pmb{x}^{(2)}\big\|},
    \label{eq:cosine}
\end{equation}
is another popular index of pattern separation. The original Hebb-Marr framework theorized pattern separation as the orthogonalization of the input patterns \cite{santoro_reassessing_2013, mcnaughton_hebb-marr_1990}. The terms \say{decorrelation} (as described by the Pearson correlation) and \say{orthogonalization}, however, are not mathematically equivalent; \cite{madar_temporal_2019} explicitly showed that pairs of spike trains can be uncorrelated without being orthogonal, or can be orthogonal without being uncorrelated. The cosine similarity has thus been used to explicitly determine whether spike trains are truly orthogonalized. %This index has also been used to measure \textit{sparsity}\task{explain} \cite{severa_combinatorial_2017}.

While the cosine similarity measures the angle between two vector representations of spike trains, the \textbf{scaling factor},
\begin{equation}
    d_\phi\left(\pmb{x}^{(1)}, \pmb{x}^{(2)}\right)
        = \frac{\big\|\pmb{x}^{(1)}\big\|}{\big\|\pmb{x}^{(2)}\big\|},
    \label{eq:sf}
\end{equation}
quantifies their difference in norm. Together, $d_\theta$ and $d_\phi$ are in theory sufficient to fully describe the similarity between two vectors in Euclidean space \cite{madar_temporal_2019}, since they focus on complementary features, angle and norm.

One implementation of the \textbf{Hamming distance}, or population distance, between two neural patterns, is given by
\begin{equation*}
    d_f = \frac{\mathrm{HD}}{2 N(1-s)},
\end{equation*}
where $\mathrm{HD}$ denotes the number of positions at which the corresponding values are different \cite{hamming_error_1950} and $s$ the sparsity. $d_f$ was used to evaluate pattern separation in several papers \cite{myers_role_2009, chavlis_dendrites_2017, chavlis_pattern_2017}. This implementation of Hamming distance is limited by its lack of consideration of the temporal aspect of a given pattern -- a cell is considered active if it fires at least once during stimulus presentation, regardless of time and frequency \cite{chavlis_dendrites_2017}. We therefore use a modified definition of the Hamming distance \cite{bird_robust_2024}, and take the average of the absolute difference at each time bin,
\begin{equation}
    d_\eta\left(\pmb{x}^{(1)}, \pmb{x}^{(2)}\right)
    = \frac{1}{N_\mathrm{neurons}\,N_\mathrm{bins}}
        \sum_{i=1}^{N_\mathrm{neurons}} \sum_{j=1}^{N_\mathrm{bins}} \left|x^{(1)}_{ij} - x^{(2)}_{ij} \right|,
    \label{eq:hamming}
\end{equation}
where permanently inactive spike trains are not removed from the ensemble and the bins in which each neuron spikes are taken into consideration.

The \textbf{SPIKE similarity} was designed to assess the dissimilarity between two spike trains \cite{kreuz_monitoring_2013}. We compute the SPIKE similarity between two neural patterns, collected from an ensemble of neurons, as the average across each neuron,
\begin{equation}
    d_S\left(T^{(1)}, T^{(2)}\right)
    = \frac{1}{N_\mathrm{neurons}} \sum_{i=1}^{N_\mathrm{neurons}}
        \delta_S\left(T^{(1)}_{i}, T^{(2)}_{i}\right).
    \label{eq:spike_similarity}
\end{equation}
The quantity
\begin{equation*}
    \delta_S\left(T_i^{(1)}, T_i^{(2)}\right)
        = 1 - \frac{1}{\tau}\int_0^\tau \dt D(t),
\end{equation*}
has been used to assess the dissimilarity between two spike trains \cite{madar_temporal_2019}. It is computed by integrating, over the observation window $[0, \tau]$, the distance between two spike trains from \cite{kreuz_monitoring_2013},
% Eq. 19 in Kreuz
\begin{equation*}
    D(t) = \frac{S\one \xisi\two + S\two \xisi\one}
        {2\left\langle \xisi^{(n)} \right\rangle^2},
    %\label{eq:spike_distance}
\end{equation*}
% Eq. 17 in Kreuz
where
\begin{equation*}
    S^{(n)}(t)
        = \frac
            {\dtp^{(n)}\xf^{(n)} + \dtf^{(n)}\xp^{(n)}}
            {\xisi^{(n)}}
    %\label{eq:spike_local_weight}
\end{equation*}
is the local weighting for the spike time differences of each spike train.

\begin{figure}[!ht]
    \centering
    \includegraphics[width=.9\linewidth]{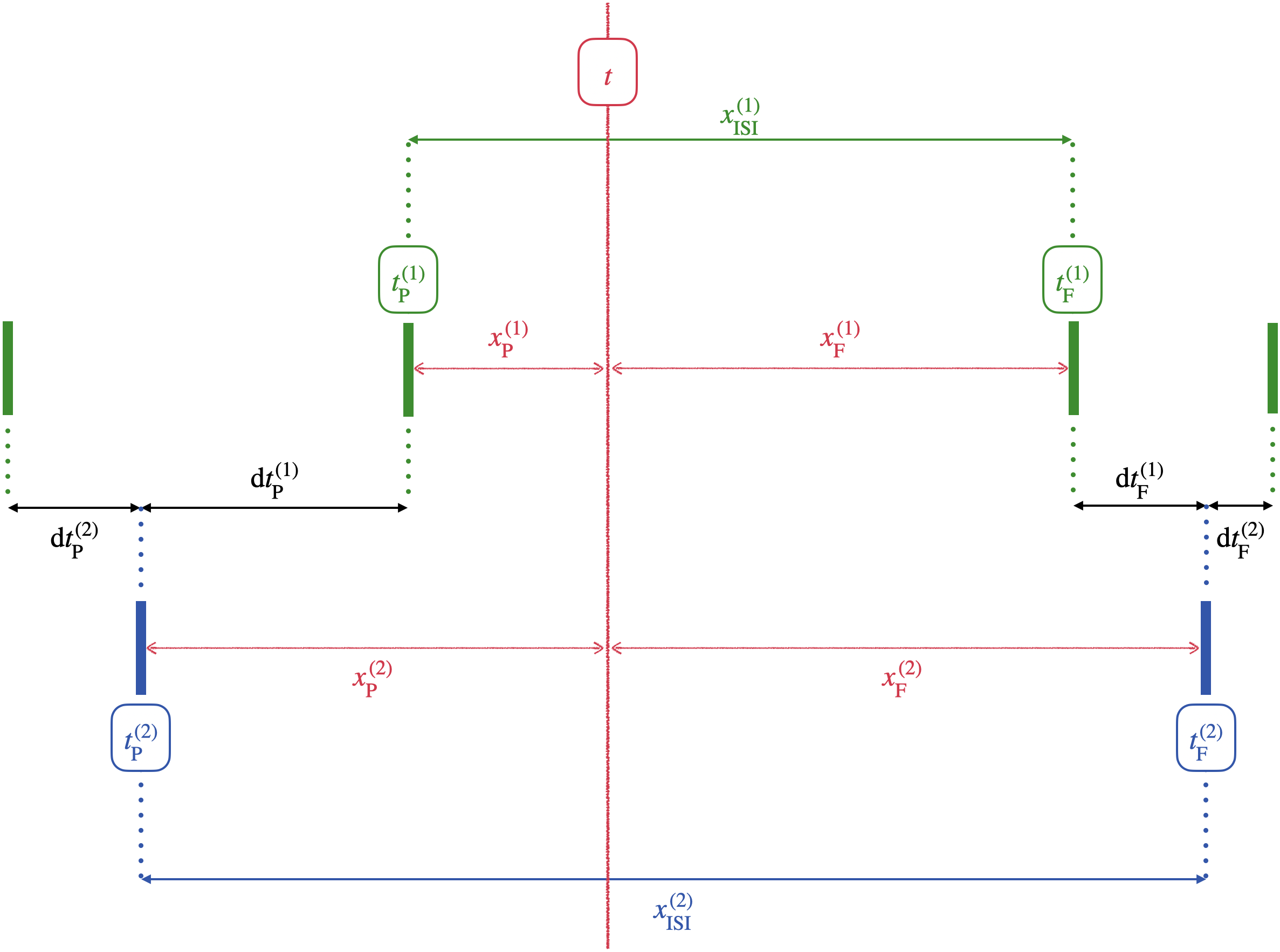}
    \caption{\textbf{Variables used to compute the SPIKE similarity.} The green and blue solid bars respectively represent the spike times of spike trains 1 and 2.}
    \label{fig:spike_params}
\end{figure}

Figure \ref{fig:spike_params} shows the variables involved in computing SPIKE similarity. For any given time $t$, the preceding (denoted with subscript P) and following (denoted with subscript F) spike times are respectively given by
% Eq. 3 and 4 in Kreuz
\begin{align*}
    \tp^{(n)}
        = \max_j \left(
            T_{ij}^{(n)}\, \middle \vert\, T_{ij}^{(n)} \le t
        \right),
    \qquad
    \tf^{(n)}
        = \min_j \left(
            T_{ij}^{(n)}\, \middle \vert\, T_{ij}^{(n)} > t
        \right).
\end{align*}
The instantaneous inter-spike interval (denoted with subscript ISI) for each neuron is
% Eq. 5 in Kreuz
$$\xisi^{(n)} = \tf^{(n)}-\tp^{(n)}.$$
The intervals to the previous and following spikes for each neuron are denoted
% Eq. 9 in Kreuz
\begin{align*}
    \xp^{(n)}=t-\tp^{(n)},
    \quad
    \xf^{(n)}=\tf^{(n)} - t.
\end{align*}
For pattern 1, the instantaneous absolute differences of the preceding and following spike times are respectively
% Eq. 16 in Kreuz
\begin{align*}
    \dtp^{(1)}
        = \min_j \left(\left|\tp^{(1)} - T_{ij}^{(2)}\right|\right),
    \qquad
    \dtf^{(1)}(t)
        = \min_j \left(\left|\tf^{(1)} - T_{ij}^{(2)}\right|\right).
\end{align*}
For pattern 2, $\dtp^{(2)}$ and $\dtf^{(2)}$ are defined analogously.

%\subsubsection{Wasserstein distance $d_{W}$}
%
%\begin{align*}
%    d_{\mathrm{W}}(\vtau_1, \vtau_2) &=
%        \int_{\min{(\tau_{11}, \tau_{21})}}^{\max{\left(\tau_{1n_1}, \tau_{2n_2}\right)}} \dt
%        \\ &\qquad
%        \Bigg |
%            \left[
%                \chi_{\tau_{1n_1}, \infty}(t)
%                + \sum_{i=1}^{n_1-1} \frac{i}{n_1} \chi_{\tau_{1i}, \tau_{1i+1}}(t)
%            \right]
%            \\ &\qquad
%            - \Bigg[
%                \chi_{\tau_{2n_2}, \infty}(t)
%                + \sum_{j=1}^{n_2-1} \frac{j}{n_2} \chi_{\tau_{2j}, \tau_{2j+1}}(t)
%            \Bigg]
%        \Bigg |
%        \NUM
%\end{align*}

\subsection{Information-Theoretic Measures}
\label{subsec:info_metrics}

In addition to common indices of pattern separation found in pattern separation research, we also evaluate three information-theoretic measures using the toolbox provided by \cite{bird_robust_2024}, who first applied these measures to neural patterns. These indices were used to show that common indices, described in the previous section, could conflate pattern separation with information loss. 

The \textbf{estimated mutual information} $\hat{d}_\mathrm{M}$ is computed based on the modified Kozachenko-Leonenko estimator, adopted from \cite{houghton_calculating_2019}. Briefly, patterns $T^{(1)}$ and $T^{(2)}$ are each divided into $N$ periods of equal size. The pairwise spike train distances between each period of each spike train are computed using the Wasserstein distance $\delta$ (see \cite{dobrushin_prescribing_1970, sihn_spike_2019, bird_robust_2024}). This produces a set of pairwise distances between periods of $T^{(1)}$ and $T^{(2)}$. A biased estimator of the mutual information between $T^{(1)}$ and $T^{(2)}$, in terms of an integer smoothing parameter $1 \le h < N$, is given by
\begin{equation}
    \hat{d}_M\left(T^{(1)}, T^{(2)}\right)
        = \frac{1}{N} \sum_{i=1}^{N}
            \log \frac{N \left|
                \mathcal{C}\left(T^{(1)}_{i}, T^{(2)}_{i}\right)
            \right|}
            {h^2},
    \label{eq:estimated_mi}
\end{equation}
where $\left|\mathcal{C}\left(T^{(1)}_{i}, T^{(2)}_{i}\right)\right|$ is number of pairs $\left(T^{(1)}_{j}, T^{(2)}_{j}\right)$ such that both $\delta\left(T^{(1)}_{i}, T^{(1)}_{j}\right) < \delta_h\left(T^{(1)}_{i}\right)$ and $\delta\left(T^{(2)}_{i}, T^{(2)}_{j}\right) < \delta_h\left(T^{(2)}_{i}\right)$, where $\delta_h\left(T_{i}\right)$ is the distance of $T_{i}$ to its $h^{\mathrm{th}}$ nearest neighbour amongst all segments in $T$.

%\citet{bird_robust_2024} used the Wasserstein metric \task{def} to compute distance, whereas \citet{houghton_calculating_2019}, who originally proposed this method, used the van Rossum metric:
%\begin{align*}
%    d_\mathrm{R}(\vx_1, \vx_2)
%        &= \sum_{i, j} e^{-|x_{1i} - x_{1j}| / \tau}
%            + \sum_{i, j} e^{-|x_{2i} - x_{2j}| / \tau}
%        %\\& \qquad
%        - 2 \sum_{i, j} e^{-|-x_{1i} - x_{2i}| / \tau}.
%\end{align*}

The \textbf{transfer entropy} \cite{schreiber_measuring_2000} from one pattern $T^{(1)}$ to another $T^{(2)}$ is the mutual information between $T^{(2)}$ at the current time $\tau$ and the history of $T^{(1)}$ conditioned on the history of $T^{(2)}$,
\begin{equation}
    d_\mathrm{T}\left[T^{(1)}: T^{(2)}\right]
        = H\left(T^{(2)}_{\tau} \middle| T^{(2)}_{t<\tau}\right)
            - H\left(T^{(2)}_{\tau} \middle| T^{(2)}_{t<\tau}, T^{(1)}_{t<\tau}\right),
    \label{eq:sparse_te}
\end{equation}
Unlike other measures, $d_\mathrm{T}$ is directional and asymmetrical (whose arguments are thus denoted using square brackets and colon $[ : ])$. In many circumstances, the transfer entropy requires less data to produce an accurate estimate than the mutual information \cite{treves_upward_1995, conrad_comparative_2020}.

The \textbf{relative redundancy reduction} is given by
\begin{equation}
    d_\mathrm{R}\left(T^{(1)}, T^{(2)}\right)
        = \left[ R\left(T^{(1)}\right) - R\left(T^{(2)}\right) \right]\;
            \hat{d}_M\left(T^{(1)}, T^{(2)}\right),
    \label{eq:red_red}
\end{equation}
where $R(T)$ is the redundancy of pattern $T$, quantifying the parts of a signal that may encode the same information. The redundancy $R$ is estimated by adopting from \cite{williams_nonnegative_2010} to apply to pattern $T$,
\begin{align*}
    R(T)
        = \min_{T_i} \left[ \hat{d}_M(T \backslash T_i, T_i) \right].
\end{align*}

%===============================================================================
% EVAL
%===============================================================================

\section{An Evaluation of Existing Pattern Separation Indices}
\label{sec:evaluation}

\begin{table*}[!ht]
    \centering
    \caption{
        \textbf{Existing indices of pattern separation evaluated in this study.}
    }
    \label{tab:existing_indices}
    \renewcommand{\arraystretch}{2}
    \begin{tabular}{ l l c }
        Symbol & Description & Equation \\
        \hline \hline
        $d_\rho\left(\pmb{x}^{(1)}, \pmb{x}^{(2)}\right)$ & Pearson correlation & \ref{eq:pearson} \\
        $d_\theta\left(\pmb{x}^{(1)}, \pmb{x}^{(2)}\right)$ & Cosine similarity & \ref{eq:cosine} \\
        $d_\phi\left(\pmb{x}^{(1)}, \pmb{x}^{(2)}\right)$ & Scaling factor & \ref{eq:sf} \\
        $d_\eta\left(\pmb{x}^{(1)}, \pmb{x}^{(2)}\right)$ & Hamming distance & \ref{eq:hamming} \\
        %$d_W\left(T^{(1)}, T^{(2)}\right)$ & Wasserstein distance & \ref{eq:wass} \\
        %$d_R\left(\pmb{x}^{(1)}, \pmb{x}^{(2)}\right)$ & van Rossum metric & \ref{eq:van_rossum} \\
        $d_S\left(T^{(1)}, T^{(2)}\right)$ & SPIKE similarity & \ref{eq:spike_similarity} \\
        \hline
        $\hat{d}_M\left(T^{(1)}, T^{(2)}\right)$ & Estimated mutual information & \ref{eq:estimated_mi} \\
        $d_T\left[T^{(1)}: T^{(2)}\right]$ & Transfer entropy & \ref{eq:sparse_te} \\
        $d_R\left(T^{(1)}, T^{(2)}\right)$ & Relative redundancy reduction & \ref{eq:red_red} \\
        \hline
    \end{tabular}
\end{table*}

Table \ref{tab:existing_indices} shows the indices compared in this study, which are described in greater detail in the previous section. These existing indices were computed on two sample patterns, one drawn from the control distribution and the other from the comparison distribution, which were described in Section \ref{sec:diverge_patterns}. Recall from Section \ref{sec:pat_rep} that an index $d$ operates one of two representations of neural patterns: binarized vectors and undiscretized spike times.

For each measurement of an index $d$, vector $\pmb{x}^{(1)}$ was generated from control the control distribution, $\pmb{x}^{(2)}$ from the comparison distribution. Recall that each of these distributions specifies a unique set of $\{q_{00}, q_{10}, q_{01}, q_{11}\}$, which can then be used as event probabilities for a categorical distribution, from which we sample $N_\mathrm{bins}$ times to construct a neural pattern $\pmb{x}$. For indices that operate on undiscretized patterns, $\pmb{x}^{(1)}$ and $\pmb{x}^{(2)}$ were converted to $T^{(1)}$ and $T^{(2)}$ by dividing a 0 to 16ms observation period into $N_\mathrm{bins}$ bins, and adding time stamps that correspond to active bins into ordered sets. Each existing index was measured on $N_\mathrm{trials}=10$ control and comparison sample pattern pairs then averaged. Pearson correlation, cosine similarity, scaling factor, and SPIKE similarity were reviewed by \cite{madar_temporal_2019} and implemented in this study using custom Python scripts. Estimated mutual information, estimated redundancy reduction, and transfer entropy were proposed by \cite{bird_robust_2024} and computed using their Matlab toolbox. The values of $N_\mathrm{bins}$ and $N_\mathrm{trials}$ were chosen to be large enough to decrease the standard error of the mean, yet still small enough to be computationally tractable.

\subsection{Pattern Separation via Dissimilar Within-Ensemble Correlation}
\label{sec:2_neuron_diff_th}

\begin{figure}[h!]
    \centering
    \begin{subfigure}[b]{.32\linewidth}
        \raggedleft
        \includegraphics[width=1\linewidth]{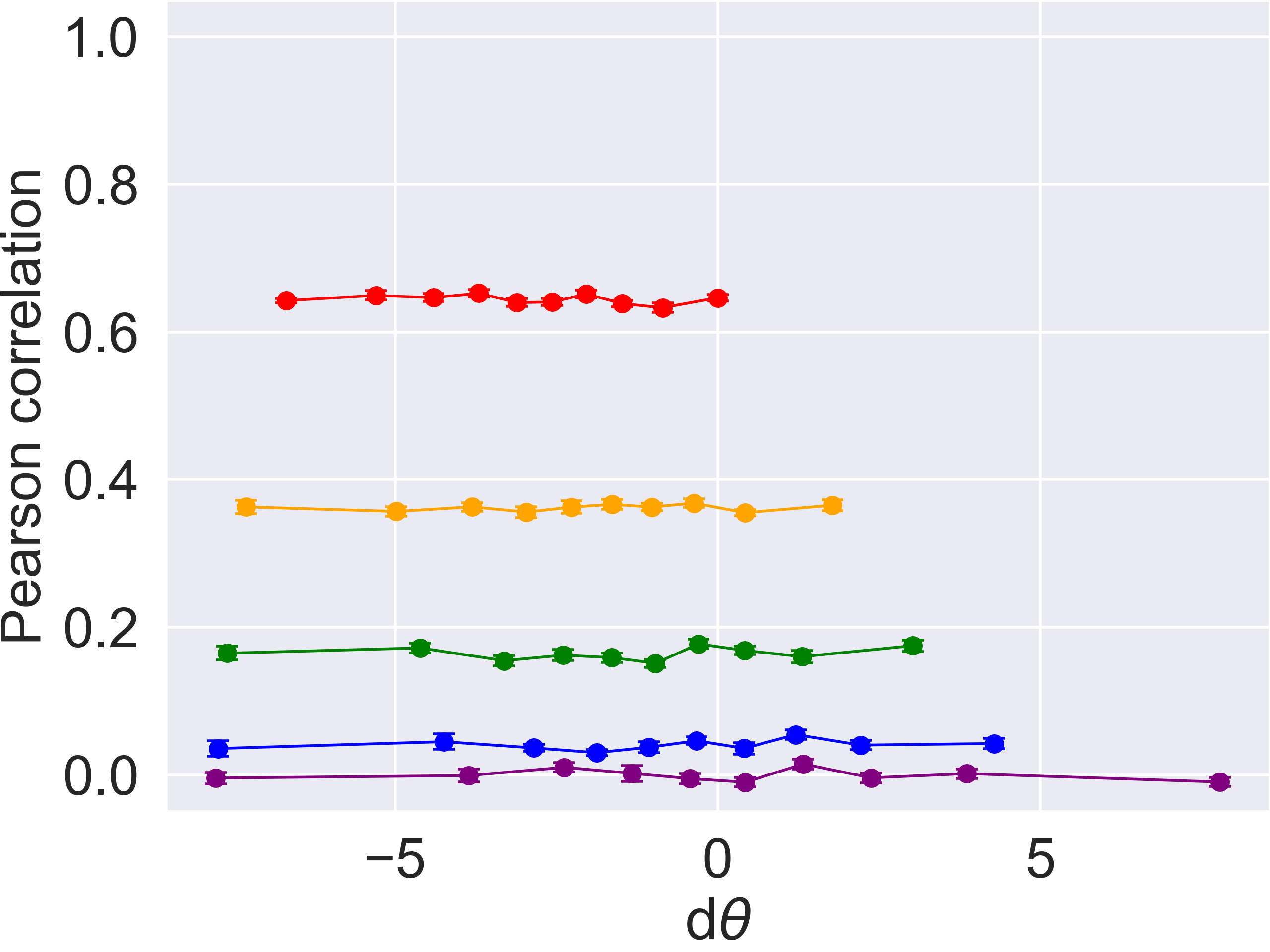}
    \end{subfigure}
    \begin{subfigure}[b]{.32\linewidth}
        \raggedleft
        \includegraphics[width=1\linewidth]{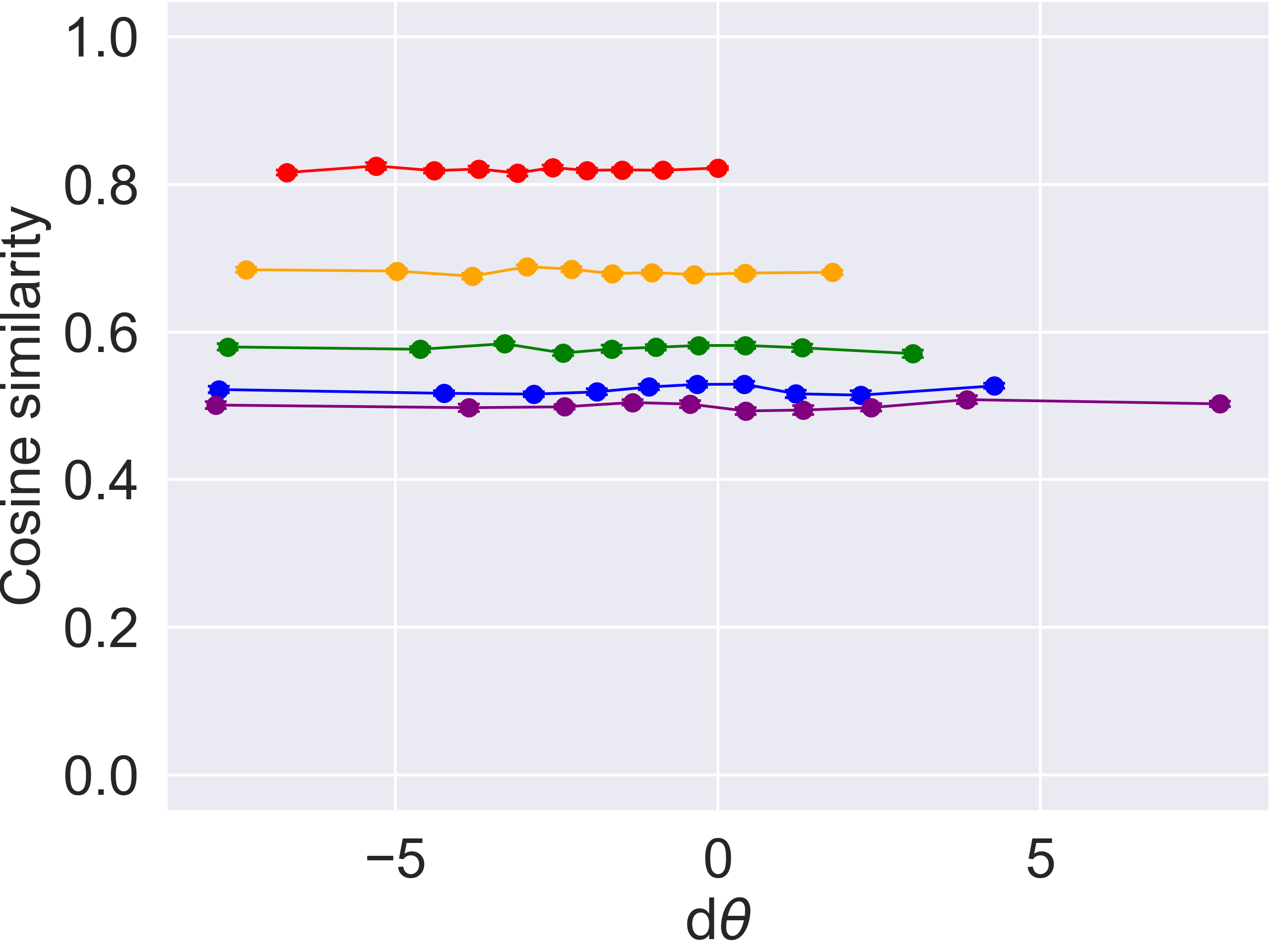}
    \end{subfigure}
    \begin{subfigure}[b]{.32\linewidth}
        \raggedleft
        \includegraphics[width=1\linewidth]{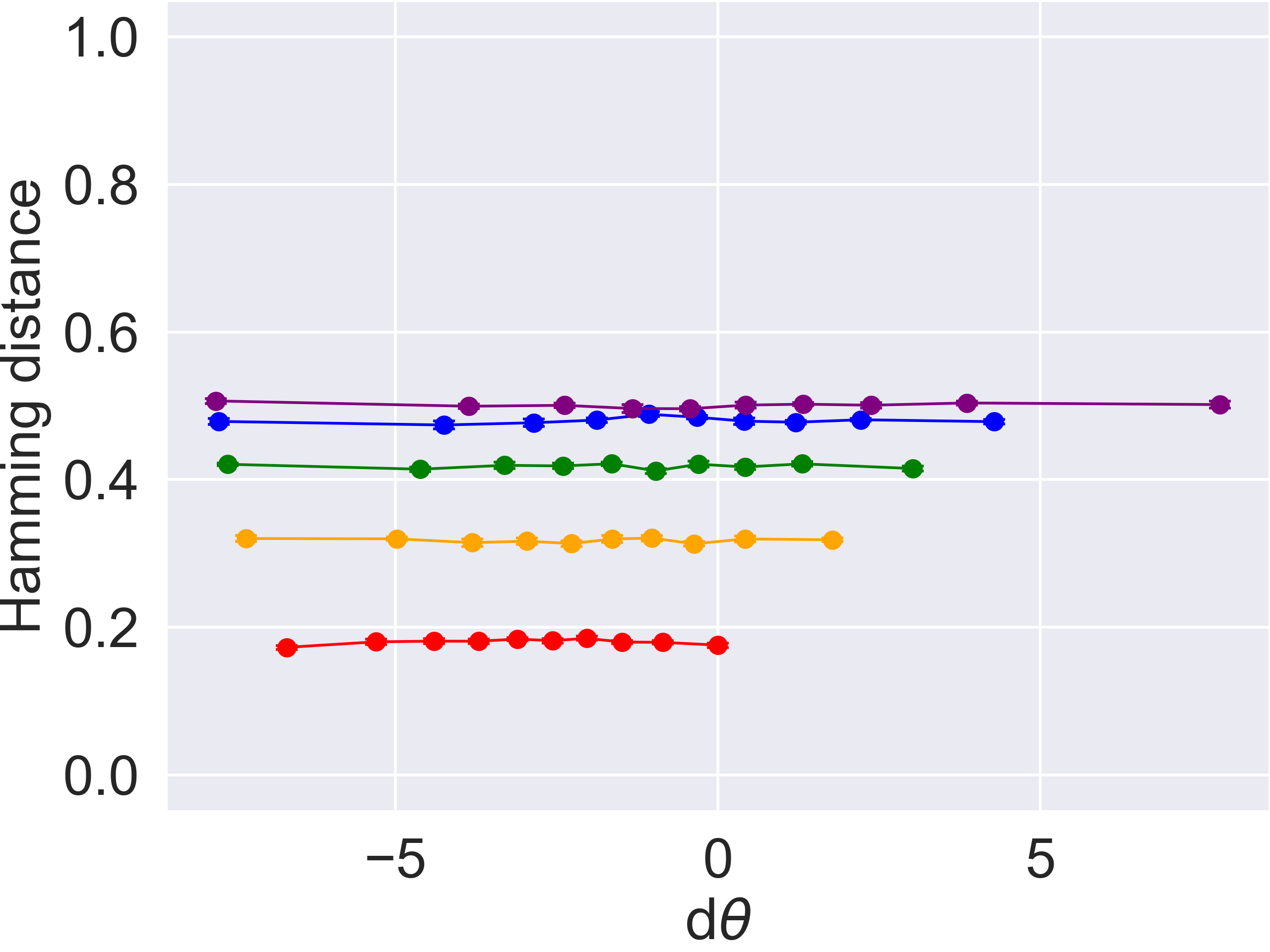}
    \end{subfigure}
    \begin{subfigure}[b]{.32\linewidth}
        \raggedleft
        \includegraphics[width=1\linewidth]{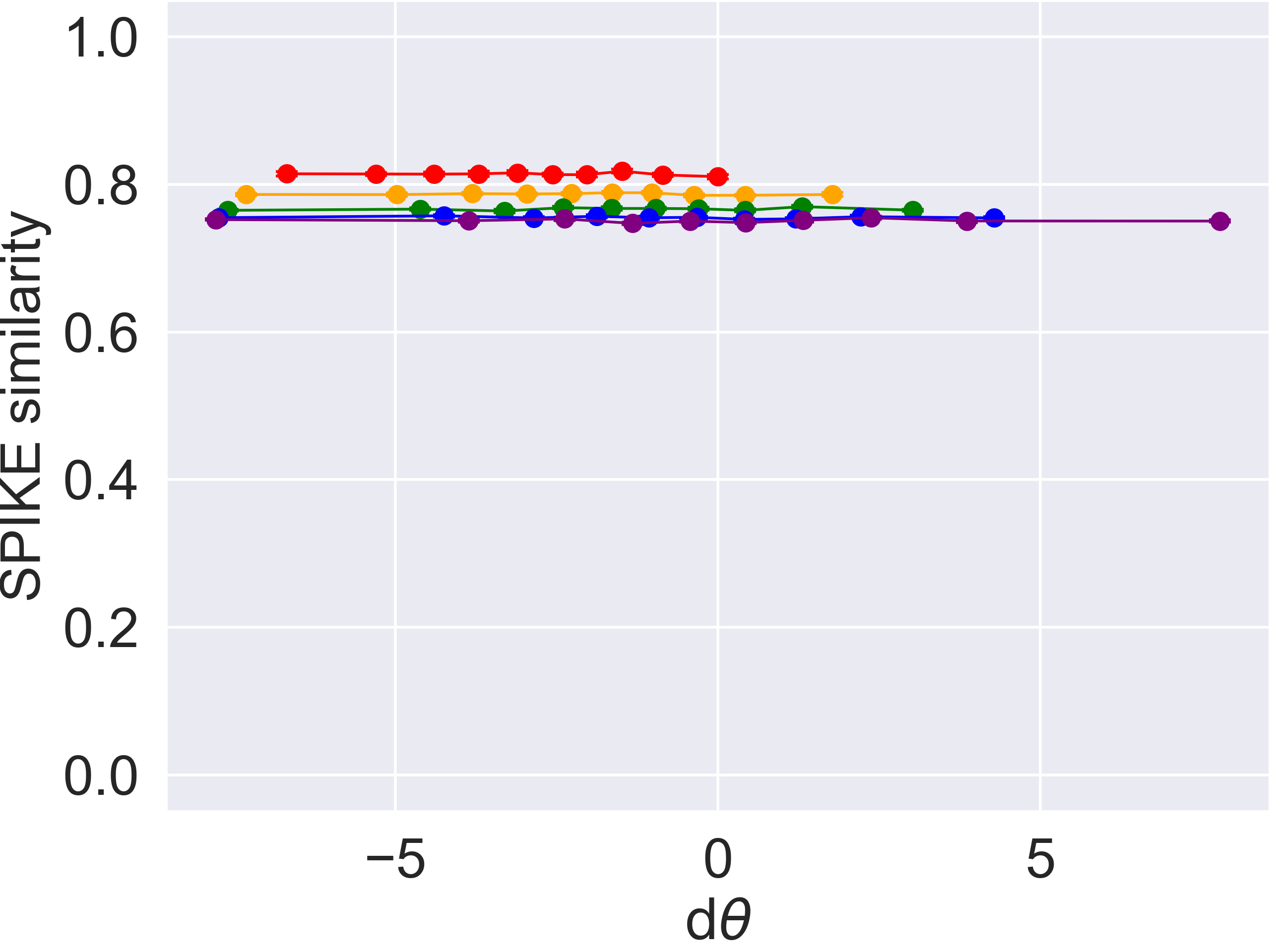}
    \end{subfigure}
    \begin{subfigure}[b]{.32\linewidth}
        \raggedleft
        \includegraphics[width=1\linewidth]{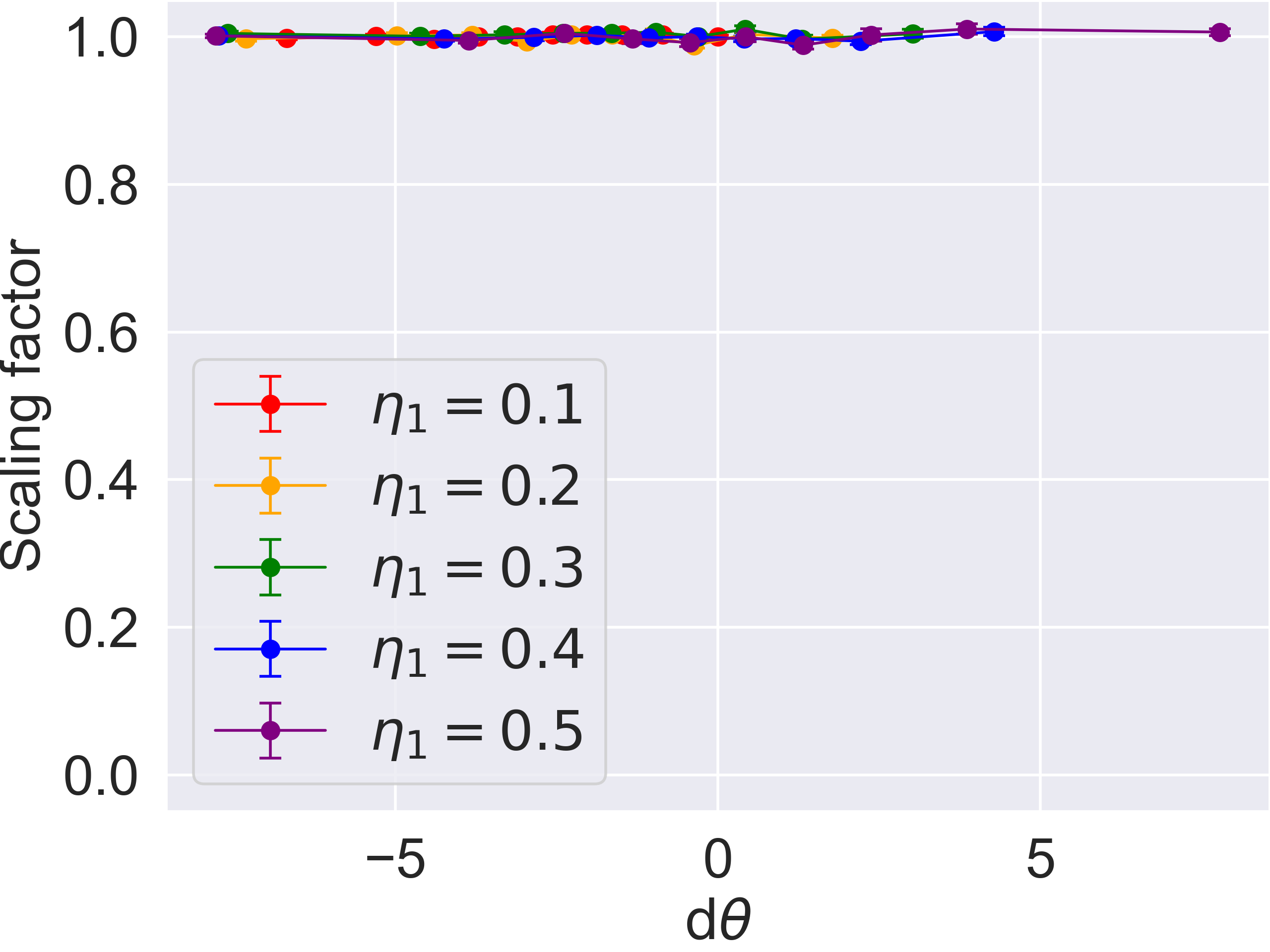}
    \end{subfigure}
    \\
    \begin{subfigure}[b]{.32\linewidth}
        \raggedleft
        \includegraphics[width=1\linewidth]{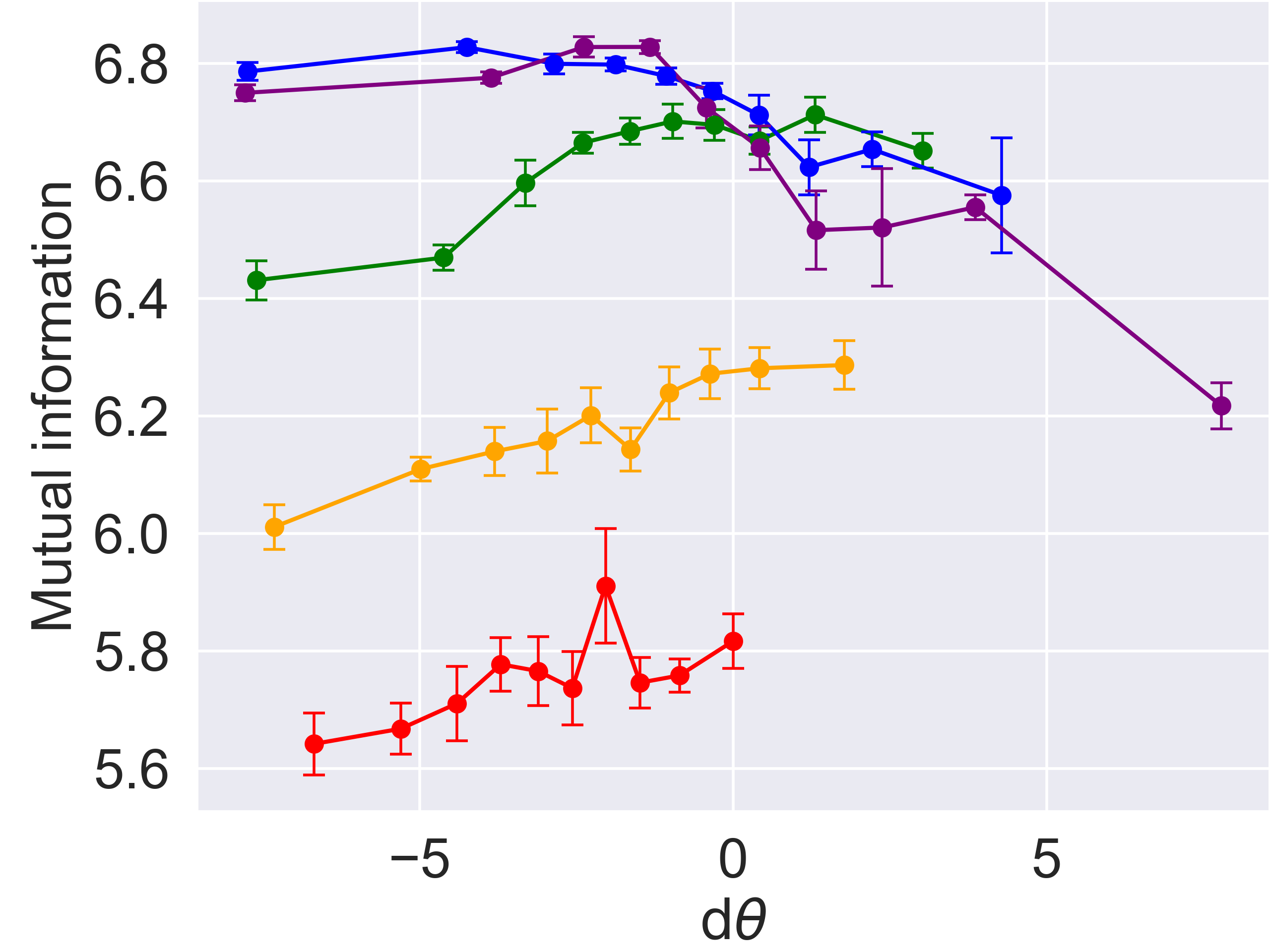}
    \end{subfigure}
    \begin{subfigure}[b]{.32\linewidth}
        \raggedleft
        \includegraphics[width=1\linewidth]{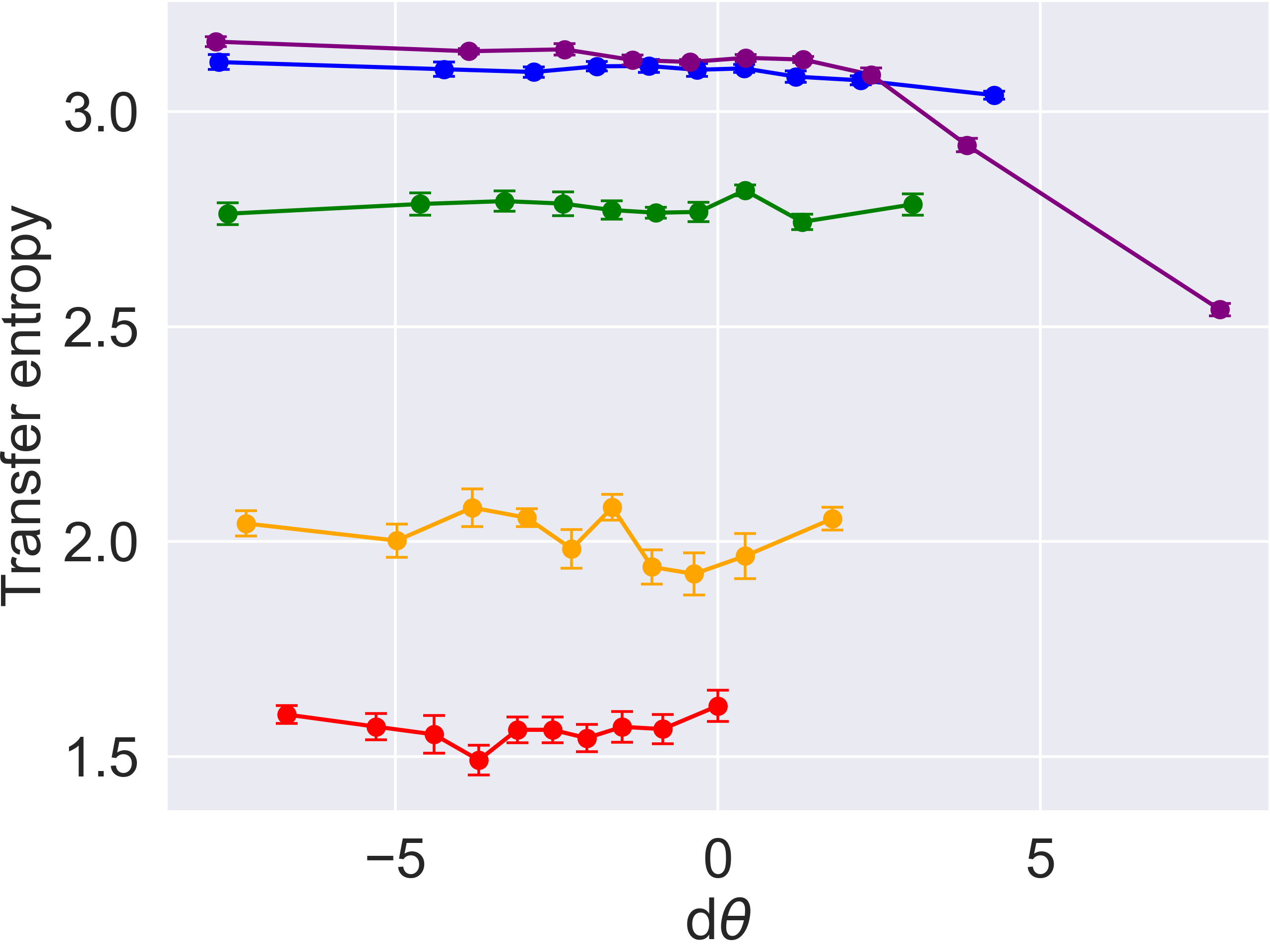}
    \end{subfigure}
    \begin{subfigure}[b]{.32\linewidth}
        \raggedleft
        \includegraphics[width=1\linewidth]{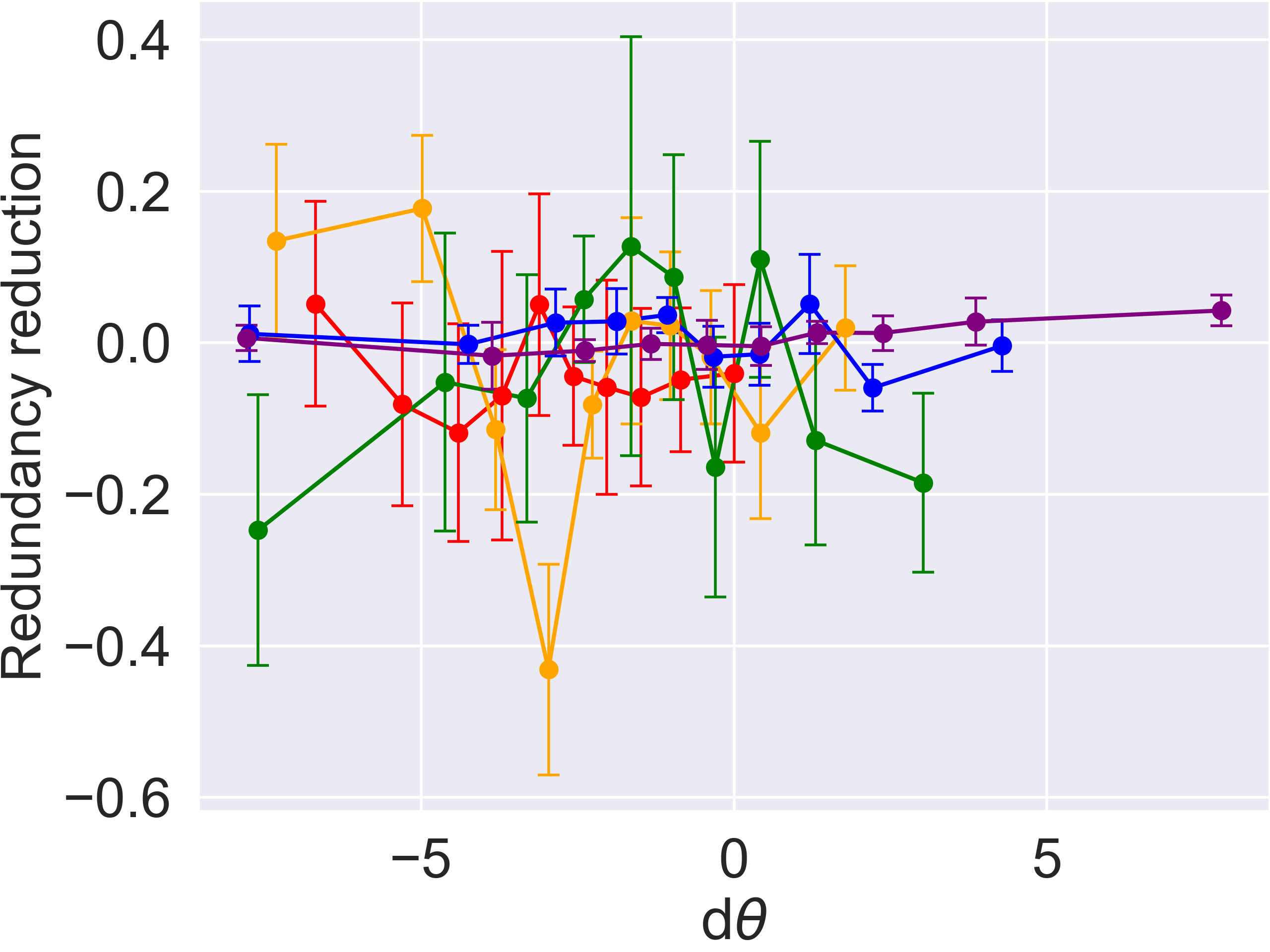}
    \end{subfigure}
    \caption{
        \textbf{Measurements of existing pattern separation indices versus change in the rate of coincidental firing}. Indices were applied to spike trains generated from $p(x; \eta_1, \eta_2, 0)$ and $p(x; \eta_1, \eta_2, \mathrm{d} \theta)$, over different values of $\eta_1$ and $\eta_2$ shown by different coloured plots (with the constraint $\eta_1 + \eta_2 = 1$), and averaged over 10 samples. The error bars show the standard error of the mean.
    }
    \label{fig:indices_vs_theta}
\end{figure}

Figure \ref{fig:indices_vs_theta} shows the average values of existing indices computed on sampled pattern pairs drawn from control and comparison distributions, $p(\pmb{x}; \eta_1, \eta_2, 0)$ and $p(\pmb{x}; \eta_1, \eta_2, \mathrm{d}\theta)$, described in Section \ref{sec:diverge_theta}. Note that these plots are not symmetrical around $\D\theta=0$, since there is an upper limit for how much the two neurons can be correlated, given that $\eta_2 \le \eta_1$. Recall from Equation \ref{eq:sq_dist_theta} that the distance between these distributions is proportional to $\mathrm{d}\theta$. As such, an appropriate index of pattern separation should be sensitive to the increase of dissimilarity between patterns from $p(\pmb{x}; \eta_1, \eta_2, 0)$ and $p(\pmb{x}; \eta_1, \eta_2, \mathrm{d}\theta)$, as the magnitude of $\mathrm{d}\theta$ grew larger.

However, Figure \ref{fig:indices_vs_theta} shows that, with the exception of mutual information and transfer entropy, existing indices were insensitive to $\mathrm{d}\theta$. No index among the Pearson correlation, cosine similarity, scaling factor, Hamming distance, and SPIKE similarity displayed changes with $\mathrm{d}\theta$ at all. The lack of change in scaling factor is most unsurprising, since this index is only designed to measure the norm of a vector; in this case, the norm of a pattern is equivalent to the sum of the number of times each neuron fires, which remains fixed throughout the experiment. There does not appear to be a change in transfer entropy, from systems whose two neurons fire independently, to those with anti-correlated neurons (i.e. when $\mathrm{d}\theta < 0$); only when the two neurons became strongly correlated (as $\mathrm{d}\theta$ became sufficiently large when $\eta_1 = \eta_2 = 0.5$) does transfer entropy decrease. The amount of redundancy reduction appeared to only vary insignificantly around 0, not matter what the value of $\D\theta$ was.

Mutual information was the only index that varied with $\D \theta$, though the trend appears to be non-monotonic. As the systems became more and more anti-correlated (i.e. as $\D\theta < 0$ grew smaller), the mutual information between patterns where the neurons' firing rates were uneven (i.e. when $|\eta_1-\eta_2|$ is large) decreased, but increased for patterns whose neurons fired more evenly. On the other hand, it always seems to decrease as the neurons become more correlated (i.e. when $\D\theta > 0$ grew larger).

\clearpage

\subsection{Patterns with Dissimilar Firing Rates in Neuron 2}
\label{sec:2_neuron_diff_e2}

\begin{figure}[h!]
    \centering
    \begin{subfigure}[b]{.32\linewidth}
        \raggedleft
        \includegraphics[width=1\linewidth]{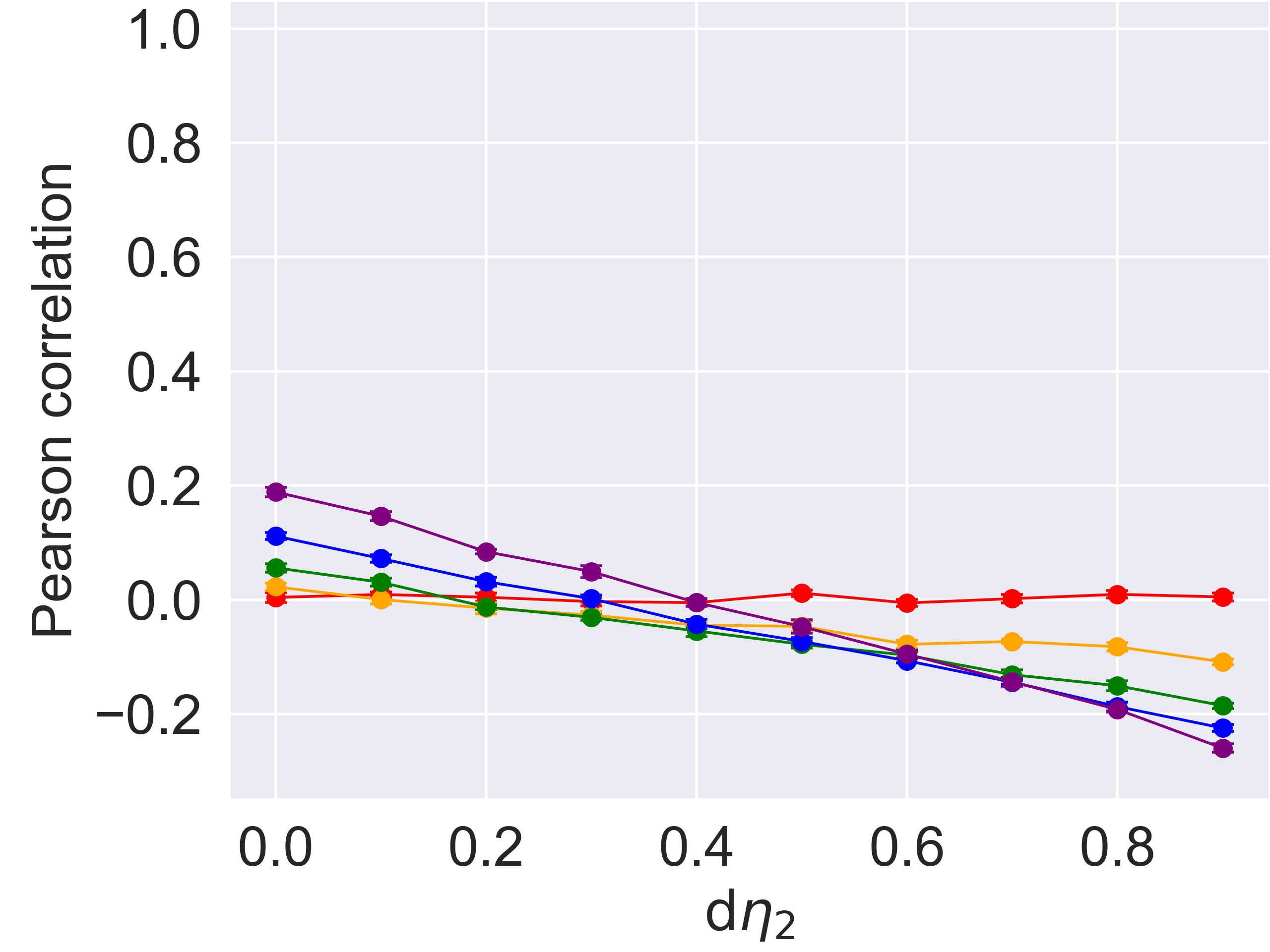}
    \end{subfigure}
    \begin{subfigure}[b]{.32\linewidth}
        \raggedleft
        \includegraphics[width=1\linewidth]{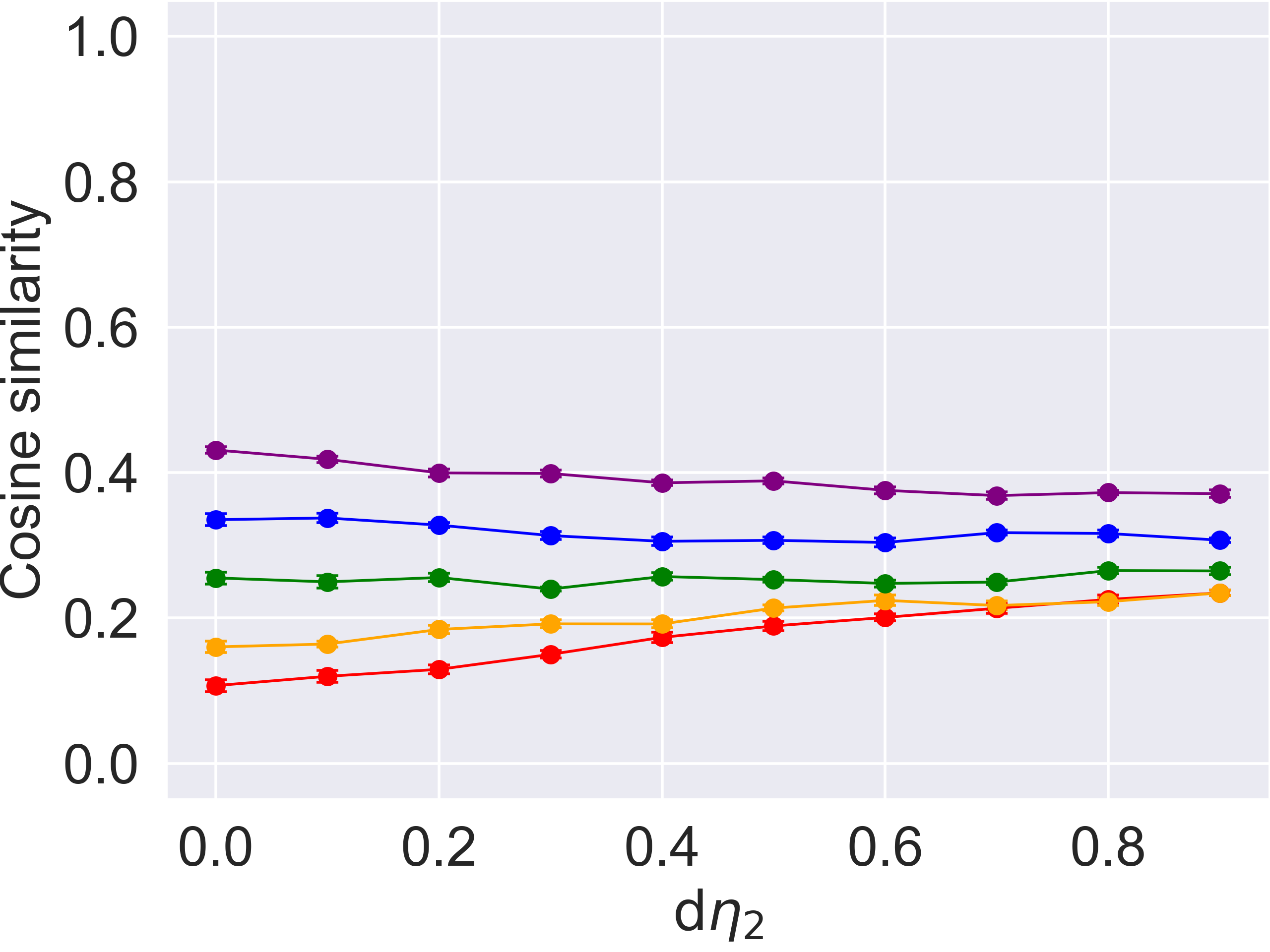}
    \end{subfigure}
    \begin{subfigure}[b]{.32\linewidth}
        \raggedleft
        \includegraphics[width=1\linewidth]{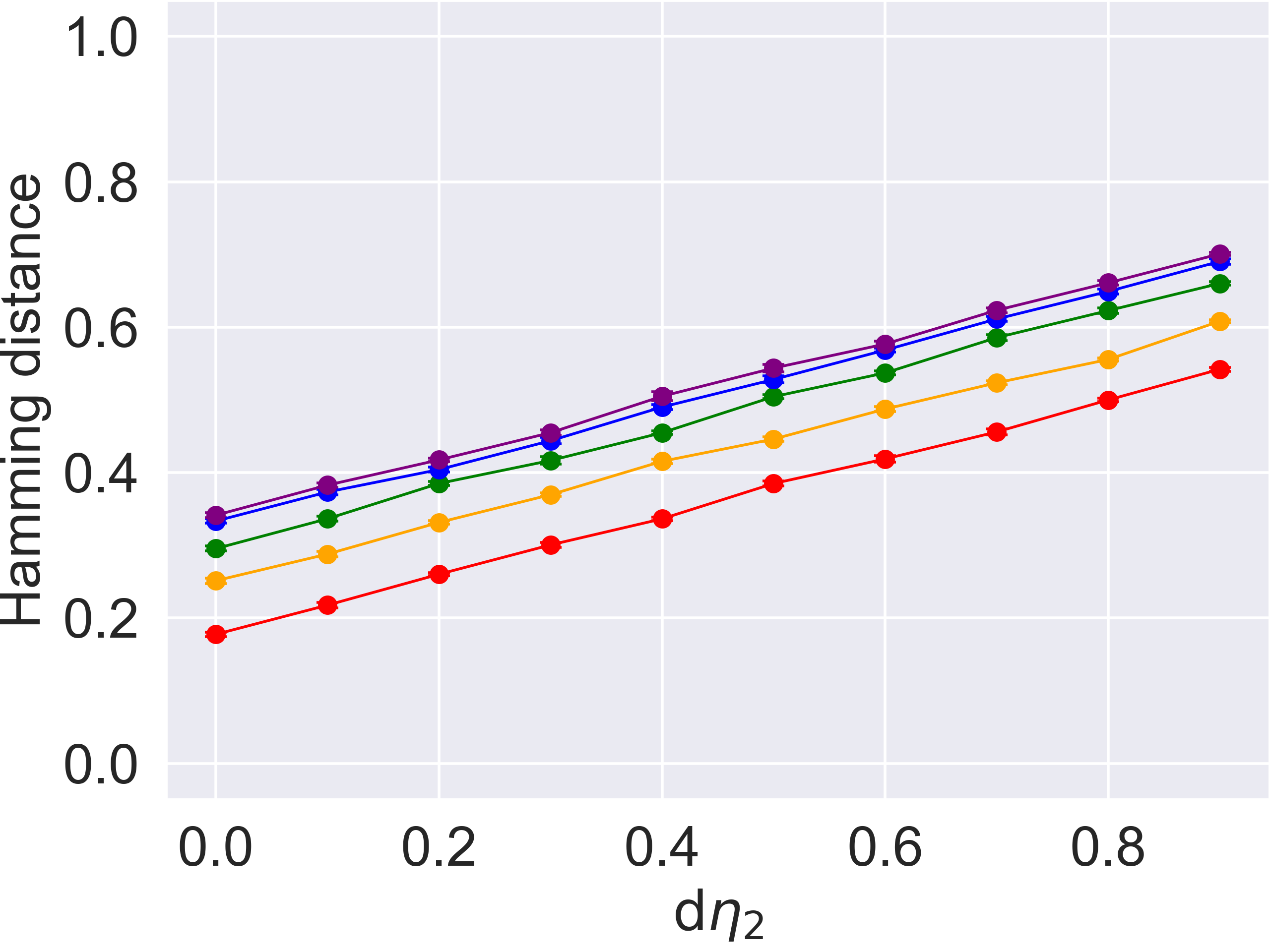}
    \end{subfigure}
    \begin{subfigure}[b]{.32\linewidth}
        \raggedleft
        \includegraphics[width=1\linewidth]{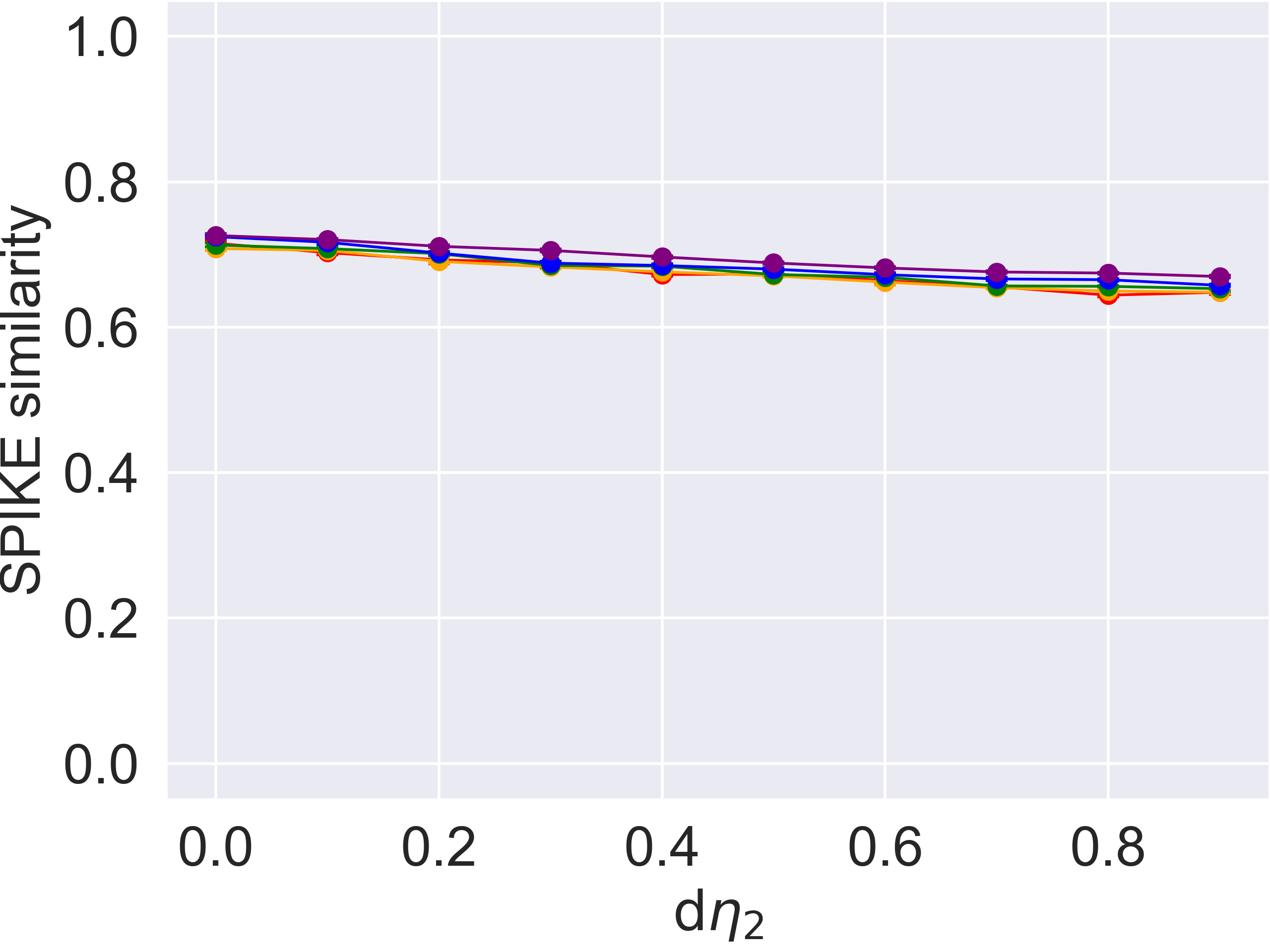}
    \end{subfigure}
    \begin{subfigure}[b]{.32\linewidth}
        \raggedleft
        \includegraphics[width=1\linewidth]{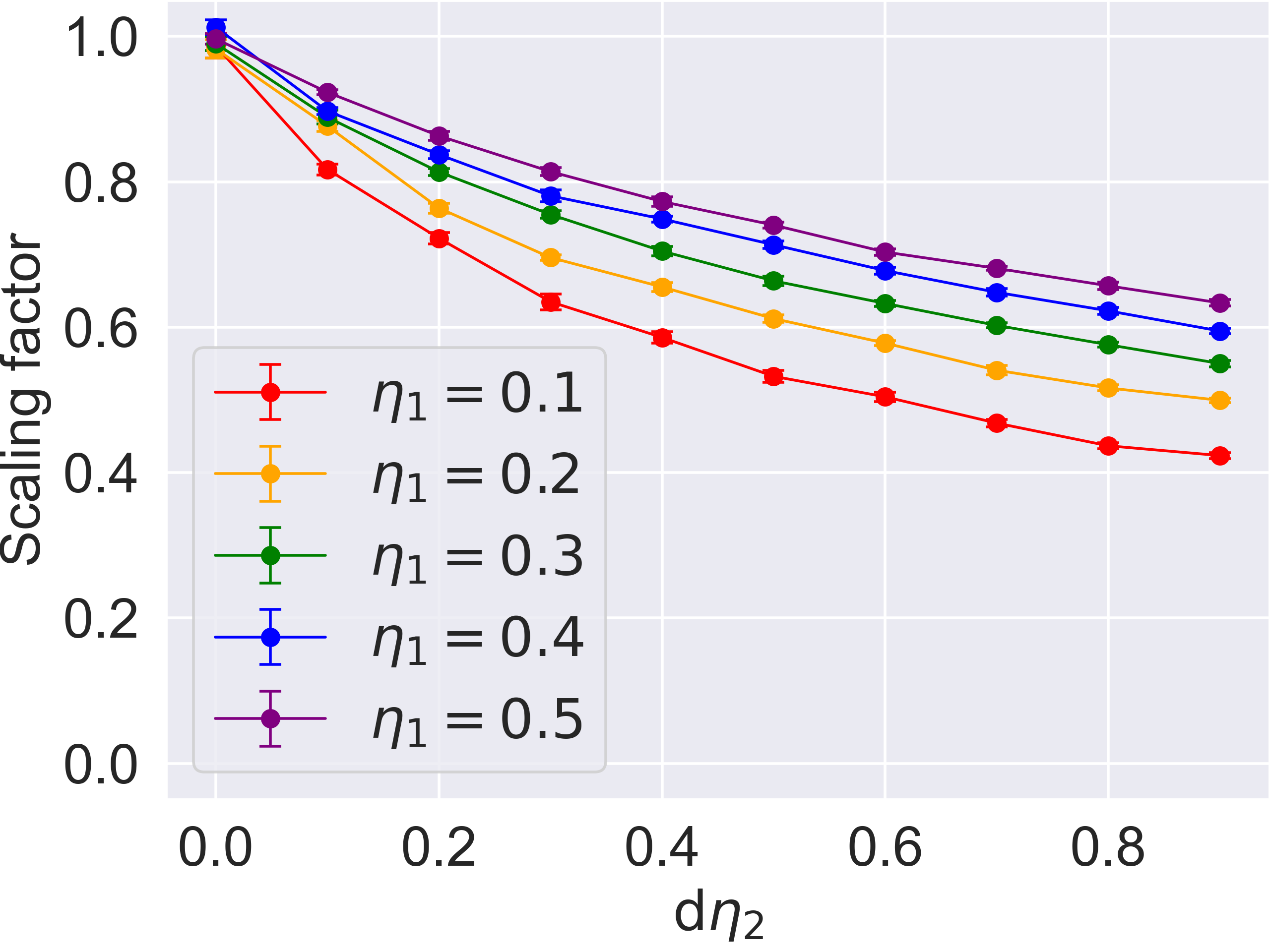}
    \end{subfigure}
    \\
    \begin{subfigure}[b]{.32\linewidth}
        \raggedleft
        \includegraphics[width=1\linewidth]{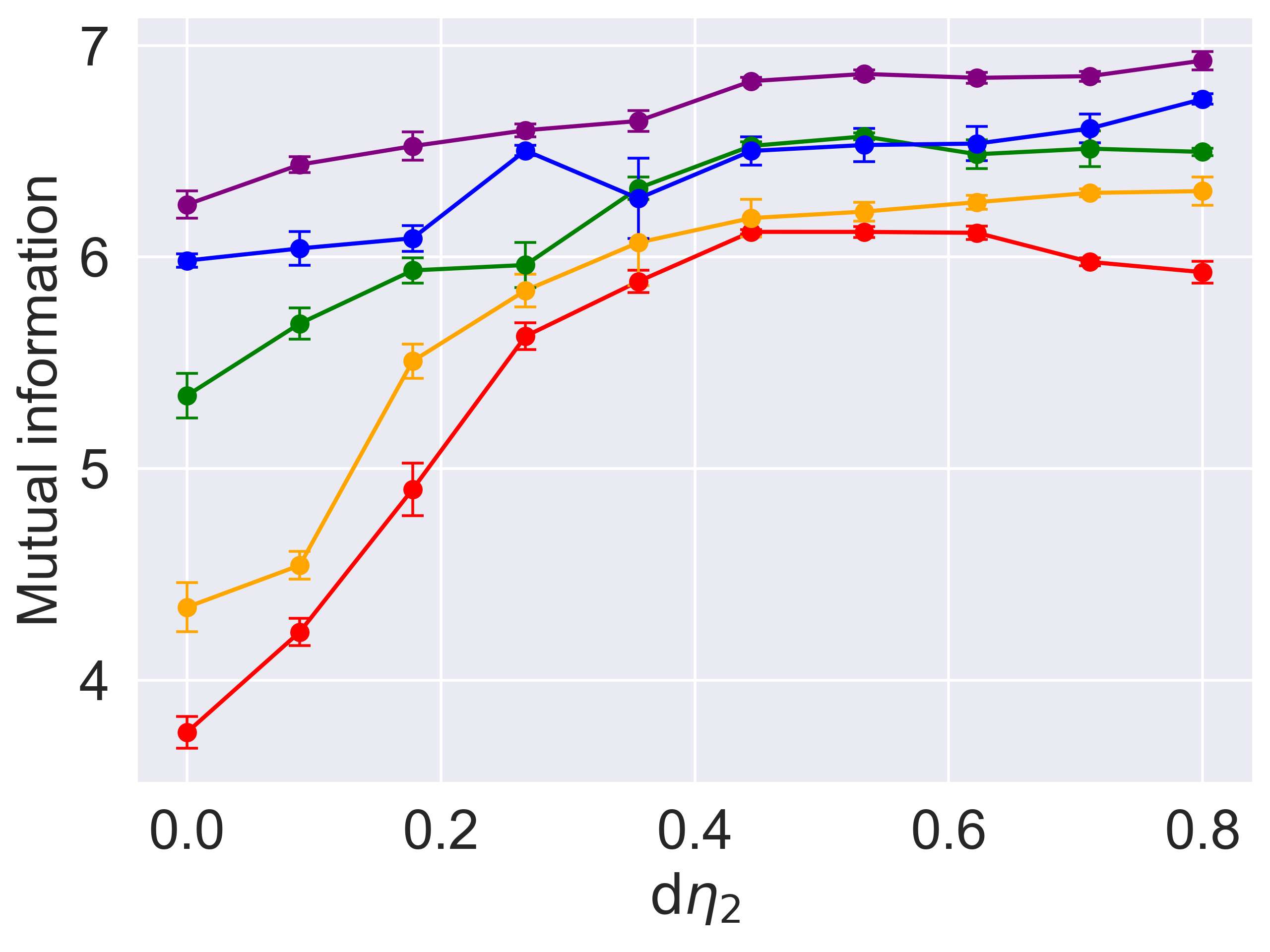}
    \end{subfigure}
    \begin{subfigure}[b]{.32\linewidth}
        \raggedleft
        \includegraphics[width=1\linewidth]{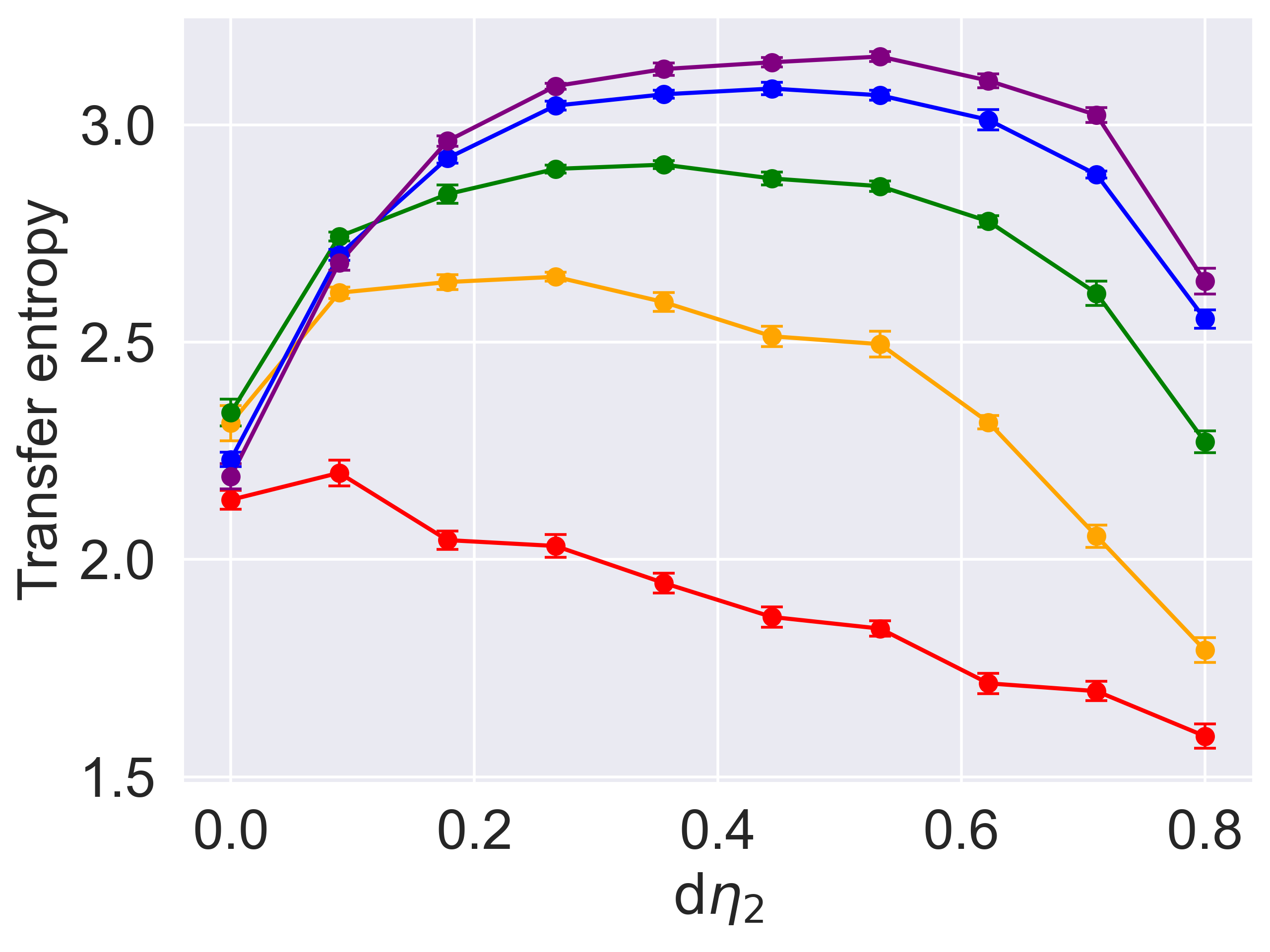}
    \end{subfigure}
    \begin{subfigure}[b]{.32\linewidth}
        \raggedleft
        \includegraphics[width=1\linewidth]{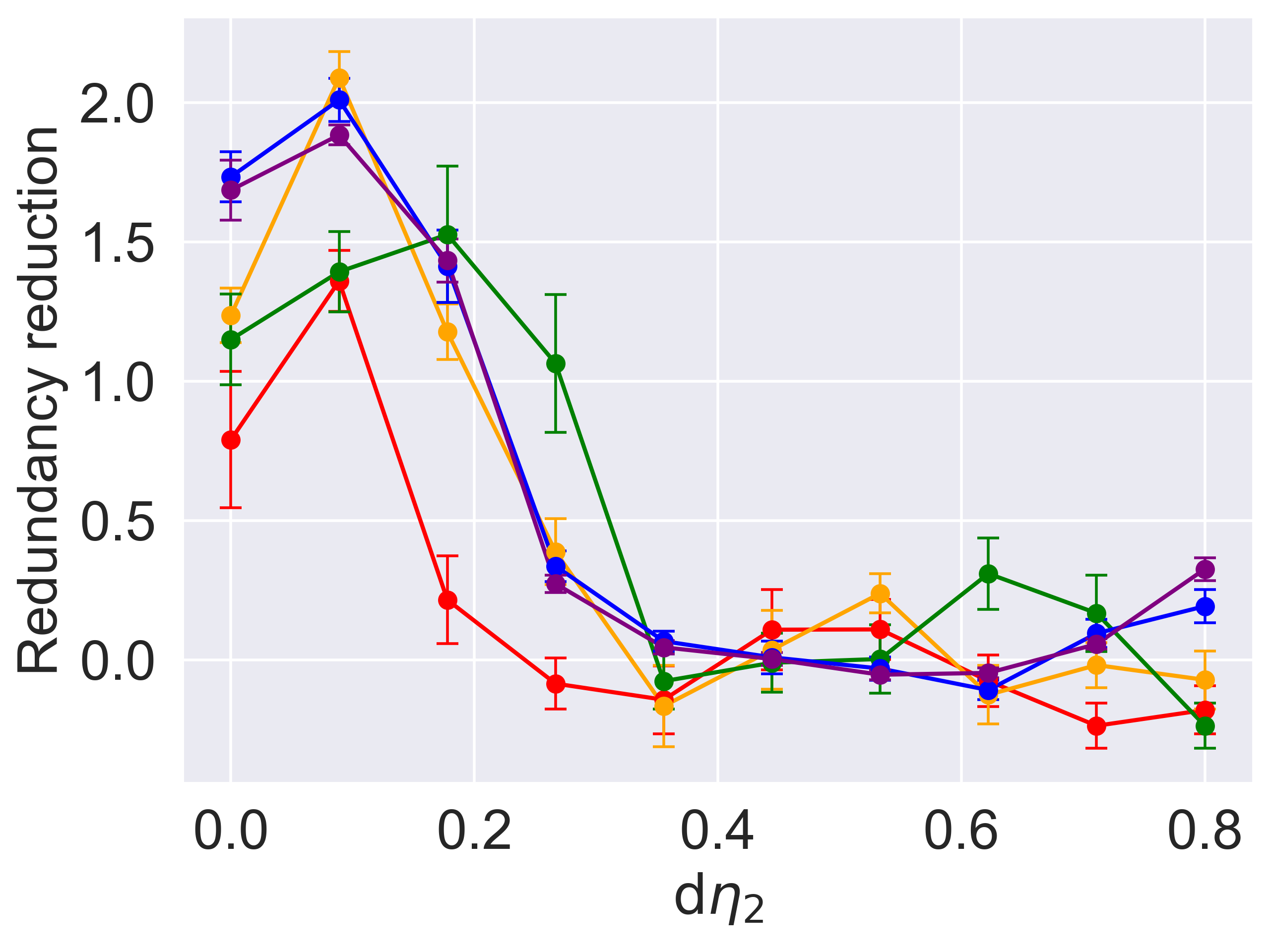}
    \end{subfigure}
    \caption{
        \textbf{Measurements of existing pattern separation indices versus change in the rate of coincidental firing}. Indices were applied to spike trains generated from $p(\pmb{x}; \eta_1, 0.1, 0)$ and $p(\pmb{x}; 0.1, 0.1 + \D\eta_2, 0)$, over different values of $\eta_1$ shown by different coloured plots, and averaged over 10 samples. The error bars show the standard error of the mean.
    }
    \label{fig:indices_vs_eta2}
\end{figure}

Figure \ref{fig:indices_vs_eta2} shows the average values of existing indices computed on sampled pattern pairs drawn from control and comparison distributions, $p(\pmb{x}; \eta_1, 0.1, 0)$ and $p(\pmb{x}; \eta_1, 0.1 + \mathrm{d}\eta_2, 0)$, where the firing rate of neuron 1 remains fixed between the two distributions, and the firing rate of neuron 2 increased (see Section \ref{sec:diverge_eta2} for more detail). Recall that the distance between these distributions is proportional to $\mathrm{d}\eta_2$. As such, an appropriate index of pattern separation should be sensitive to the increase of dissimilarity between patterns from $p(\pmb{x}; \eta_1, 0.1, 0)$ and $p(\pmb{x}; \eta_1, 0.1 + \mathrm{d}\eta_2, 0)$, as the magnitude of $\mathrm{d}\eta_2$ grew larger.

Interestingly, Pearson correlation is less sensitive to neuron 2's increase in activity if neuron 1 fires sparsely, and that the opposite is true for cosine similarity. This is another demonstration, in addition to that from \cite{madar_temporal_2019}, of how different indices could lead to different interpretations of pattern separation. 

%===============================================================================
% Section 4
%===============================================================================

\section{Discussion}
\label{sec:discussion}

We have conceptualized pattern separation in information-geometric terms, and instantiated a simple, highly controlled example using a two-neuron system with a closed-form probability law. This allowed us to test the behaviour of commonly used pattern separation measures under tightly controlled circumstances. Specifically, our simple model could generate separated patterns by either (A) changing \textit{which} neurons were active (i.e. altering each neuron's marginal firing rates), or (B) changing the correlation of activity between the two neurons.

Our work builds on previous evaluations of pattern separation indices. \cite{madar_temporal_2019} evaluated commonly used indices using recordings of input and output spike trains of single hippocampal neurons, and showed that different indices could yield different results and interpretations. Our results further support their observation. We also evaluated different pattern separation indices for specific weaknesses, much like \cite{bird_robust_2024} had done, but with a clearer mathematical formulation of pattern separation. In doing so, we showed that existing pattern separation measures, including the information-theoretic measures proposed by \cite{bird_robust_2024}, fail to recognize pattern separation when it is implemented by changing neuronal co-activation rates, rather than individual neurons' marginal firing rates. This limitation of existing measures may restrict the study of pattern separation in circuits in which pattern separation occurs via a temporal coding mechanism \cite{madar_temporal_2019, madar_pattern_2019}. 

Our information-geometric formulation of pattern separation can be extended to include more complex neural dynamics. The two-neuron model served as a simple example that allowed us to isolate the effects neuronal co-activation. Clearly, it omits many detailed behaviours of real neural systems. For example, the distribution $p(\vx; \eta_1, \eta_2, \theta)$ describes neural systems in equilibrium (i.e. the probability of neural activity is time-independent), and different families of distributions (i.e. a different statistical manifolds) could be employed to introduce temporal structure of spikes or refractoriness after a spike \cite{amari_information_2022, grun_information_2010}.

Our results encourage a search for new pattern separation indices. Neuronal co-activation, as implemented in our study, may occur due to a third neuron in a separate layer that projects into a larger layer. This neuronal circuit architecture is consistent with expansion re-coding \cite{oreilly_hippocampal_1994, cayco-gajic_re-evaluating_2019}. Additionally, increasing the level of connectivity within a neuronal population, such as dentate gyrus circuits with high degrees of mossy-fiber sprouting, may also lead to increased neuronal co-activation rates. Mossy-fiber sprouting is commonly studied in epilepsy, for which pattern separation deficits have been reported \textit{in silico} \cite{yim_intrinsic_2015}. To sufficiently study pattern separation within these systems, we argue that new measures must be developed that are capable of capturing the effects of neuronal co-activation rates on pattern separation. One such measure could, of course, be the distance between probability distributions underlying different neural patterns. For high-dimensional systems (i.e. models with many neurons and/or parameters), however, we would require efficient methods by which geodesic distances between distributions can be estimated. This requires efficiently estimating Fisher information matrices $G$, the dimension for which increases exponentially with the number of neurons even in simple models \cite{nakahara_information-geometric_2002}. However, progress in this area may help us better understand the computational nature of pattern separation, as well as neural systems that perform pattern separation.

\section*{Acknowledgments}
This research was funded by Research Nova Scotia grant number RNS-NHIG-2021-1931.

%Bibliography
\bibliographystyle{unsrt}  
\bibliography{references}

\end{document}